\def\hang{}
\def\textindent{}
\def\par{}
\def\beginparmode{\endmode
  \begingroup \def\endmode{\par\endgroup}}
\let\endmode=\par
{\obeylines\gdef\
{}}

\def\body			
  {\beginparmode}		

\def\head#1{			
  \goodbreak\vskip 0.5truein	
  {\immediate\write16{#1}
   \uppercase{#1}\par}
   \nobreak\vskip 0.25truein\nobreak}

\def\itemj{\par\hang\textindent}
\def\beginitems{
\par\medskip\bgroup\def\i##1 {\itemj{##1}}\def\ii##1 {\itemitem{##1}}
\leftskip=36pt\parskip=0pt}
\def\enditems{\par\egroup}

\def\beneathrel#1\under#2{\mathrel{\mathop{#2}\limits_{#1}}}

\def\refto#1{[#1]}		

\def\references			
  {\head{REFERENCES}		
   \beginparmode
   \frenchspacing \parindent=0pt \leftskip=1truecm
   \parskip=8pt plus 3pt \everypar{\hangindent=\parindent}}

\gdef\refis#1{\itemj{#1.\ }}			

\gdef\journal#1, #2, #3, 1#4#5#6{		
    {\sl #1~}{\bf #2} (1#4#5#6), #3 }		

\def\annp{\journal Ann. Phys. (N.Y.), }

\def\Annp{\journal Ann. Physik, }

\def\aspm{\journal Advanced Studies in Pure Mathematics, }

\def\cmp{\journal Comm. Math. Phys., }

\def\eurolett{\journal Europhysics Lett., }

\def\ijmpa{\journal Int. J. Mod. Phys. A, }

\def\ijmpb{\journal Int. J. Mod. Phys. B, }

\def\jappp{\journal J. Appl. Phys., }

\def\jphc{\journal J. Physique C, }

\def\jpI{\journal J. Physique I, }

\def\jpcoll{\journal J. Physique Coll, }

\def\jcr{\journal J. Chem. Res., }

\def\jetp{\journal Sov. Phys. JETP, }

\def\jetpl{\journal JETP Lett., }

\def\jpj{\journal J. Phys. Soc. Japan, }

\def\jmp{\journal J. Math. Phys., }

\def\jpa{\journal J. Phys. A, }

\def\jpc{\journal J. Phys. C, }

\def\jpcon{\journal J. Phys.: Condens. Matter, }

\def\ptp{\journal Prog. Theor. Phys., }

\def\jetp{\journal Sov. Phys. JETP, }

\def\jpsj{\journal J. Phys. Soc. Japan, }

\def\jsp{\journal J. Stat. Phys., }

\def\lmp{\journal Lett. Math. Phys., }

\def\lnp{\journal Lecture Notes in Physics, }

\def\mpla{\journal Mod. Phys. Lett. A, }

\def\npb{\journal Nucl. Phys. B, }

\def\physica{\journal Physica, }

\def\pla{\journal Phys. Lett. A, }

\def\plb{\journal Phys. Lett. B, }

\def\prep{\journal Physics Reports, }

\def\pra{\journal Phys. Rev. A, }

\def\prb{\journal Phys. Rev. B, }

\def\prl{\journal Phys. Rev. Lett., }

\def\prs{\journal Proc. Roy. Soc. (London) A, }

\def\pr{\journal Phys. Rev., }

\def\rmp{\journal Rev. Mod. Phys., }

\def\sjnp{\journal Sov. J. Nucl. Phys., }

\def\tmp{\journal Theor. Math. Phys., }

\def\zpb{\journal Z. Phys. B, }

\def\zp{\journal Z. Phys., }

\def\reff#1{Ref.~#1}			
\def\Reff#1{Ref.~#1}			
\def\[#1]{[\refcite{#1}]}
\def\refcite#1{{#1}}
\def\(#1){(\call{#1})}
\def\call#1{{#1}}
\def\taghead#1{}
\def\frac#1#2{{#1 \over #2}}

\def\sla{\raise.15ex\hbox{$/$}\kern-.57em}
\def\leaderfill{\leaders\hbox to 1em{\hss.\hss}\hfill}
\def\twiddle{\lower.9ex\rlap{$\kern-.1em\scriptstyle\sim$}}
\def\bigtwiddle{\lower1.ex\rlap{$\sim$}}
\def\gtwid{\mathrel{\raise.3ex\hbox{$>$\kern-.75em\lower1ex\hbox{$\sim$}}}}
\def\ltwid{\mathrel{\raise.3ex\hbox{$<$\kern-.75em\lower1ex\hbox{$\sim$}}}}
\def\square{\kern1pt\vbox{\hrule height 1.2pt\hbox{\vrule width 1.2pt\hskip
3pt
   \vbox{\vskip 6pt}\hskip 3pt\vrule width 0.6pt}\hrule height
0.6pt}\kern1pt}
\def\tdot#1{\mathord{\mathop{#1}\limits^{\kern2pt\ldots}}}

\def\pmb#1{\setbox0=\hbox{#1}%
  \kern-.025em\copy0\kern-\wd0
  \kern  .05em\copy0\kern-\wd0
  \kern-.025em\raise.0433em\box0 }

\def\refto#1{[#1]}		

\def\references			
  {\section*{References}		
   \beginparmode
   \frenchspacing \parindent=0pt \leftskip=1truecm
   \parskip=8pt plus 3pt \everypar{\hangindent=\parindent}}

\def\endreferences{\body}

\def\reff#1{Ref.~#1}			
\def\Reff#1{Ref.~#1}			
\def\[#1]{[\refcite{#1}]}
\def\refcite#1{{#1}}

\catcode`@=11
\newcount\r@fcount \r@fcount=0
\newcount\r@fcurr
\immediate\newwrite\reffile
\newif\ifr@ffile\r@ffilefalse
\def\w@rnwrite#1{\ifr@ffile\immediate\write\reffile{#1}\fi\message{#1}}

\def\writer@f#1>>{}
\def\referencefile{
  \r@ffiletrue\immediate\openout\reffile=\jobname.ref%
  \def\writer@f##1>>{\ifr@ffile\immediate\write\reffile%
    {\noexpand\refis{##1} = \csname r@fnum##1\endcsname = %
     \expandafter\expandafter\expandafter\strip@t\expandafter%
     \meaning\csname r@ftext\csname r@fnum##1\endcsname\endcsname}\fi}%
  \def\strip@t##1>>{}}

\def\citeall#1{\xdef#1##1{#1{\noexpand\refcite{##1}}}}
\def\refcite#1{\each@rg\citer@nge{#1}}	

\def\each@rg#1#2{{\let\thecsname=#1\expandafter\first@rg#2,\end,}}
\def\first@rg#1,{\thecsname{#1}\apply@rg}	
\def\apply@rg#1,{\ifx\end#1\let\next=\relax
\else,\thecsname{#1}\let\next=\apply@rg\fi\next}

\def\citer@nge#1{\citedor@nge#1-\end-}	
\def\citer@ngeat#1\end-{#1}
\def\citedor@nge#1-#2-{\ifx\end#2\r@featspace#1 
  \else\citel@@p{#1}{#2}\citer@ngeat\fi}	
\def\citel@@p#1#2{\ifnum#1>#2{\errmessage{Reference range #1-#2\space is
bad.}
    \errhelp{If you cite a series of references by the notation M-N, then M
and
    N must be integers, and N must be greater than or equal to M.}}\else%
 {\count0=#1\count1=#2\advance\count1
by1\relax\expandafter\r@fcite\the\count0,%
  \loop\advance\count0 by1\relax
    \ifnum\count0<\count1,\expandafter\r@fcite\the\count0,%
  \repeat}\fi}

\def\r@featspace#1#2 {\r@fcite#1#2,}	
\def\r@fcite#1,{\ifuncit@d{#1}
    \newr@f{#1}%
    \expandafter\gdef\csname r@ftext\number\r@fcount\endcsname%
                     {\message{Reference #1 to be supplied.}%
                      \writer@f#1>>#1 to be supplied.\par}%
 \fi%
 \csname r@fnum#1\endcsname}
\def\ifuncit@d#1{\expandafter\ifx\csname r@fnum#1\endcsname\relax}%
\def\newr@f#1{\global\advance\r@fcount by1%
    \expandafter\xdef\csname r@fnum#1\endcsname{\number\r@fcount}}

\let\r@fis=\refis			
\def\refis#1#2#3\par{\ifuncit@d{#1}
   \newr@f{#1}%
   \w@rnwrite{Reference #1=\number\r@fcount\space is not cited up to
now.}\fi%
  \expandafter\gdef\csname r@ftext\csname r@fnum#1\endcsname\endcsname%
  {\writer@f#1>>#2#3\par}}

\def\ignoreuncited{
   \def\refis##1##2##3\par{\ifuncit@d{##1}%
     \else\expandafter\gdef\csname r@ftext\csname
r@fnum##1\endcsname\endcsname%
     {\writer@f##1>>##2##3\par}\fi}}

\def\r@ferr{\endreferences\errmessage{I was expecting to see
\noexpand\endreferences before now;  I have inserted it here.}}
\let\r@ferences=\references
\def\references{\r@ferences\def\endmode{\r@ferr\par\endgroup}}

\let\endr@ferences=\endreferences
\def\endreferences{\r@fcurr=0
  {\loop\ifnum\r@fcurr<\r@fcount
    \advance\r@fcurr by
1\relax\expandafter\r@fis\expandafter{\number\r@fcurr}%
    \csname r@ftext\number\r@fcurr\endcsname%
  \repeat}\gdef\r@ferr{}\endr@ferences}

\let\r@fend=\endpaper\gdef\endpaper{\ifr@ffile
\immediate\write16{Cross References written on []\jobname.REF.}\fi\r@fend}

\catcode`@=12

\citeall\refto		
\citeall\reff		%
\citeall\Reff		%

\ignoreuncited
\newcommand\be{\begin{equation}}
\newcommand\ee{\end{equation}}
\def\bea{\begin{eqnarray}}
\def\eea{\end{eqnarray}}
\def\N{{\frak L}}
\def\A{\frak A}
\def\bA{\overline{{\frak A}}}
\def\B{{\frak A}_\N}
\def\bB{\overline{{\frak A}}_\N}
\def\tB{\widetilde{\frak A}_\N}
\def\tbB{\widetilde{\overline{{\frak A}}}_\N}
\def\a{\frak a}
\def\ba{\overline{{\frak a}}}
\def\b{{\frak a}_\N}
\def\bb{\overline{{\frak a}}_\N}
\def\tb{\widetilde{{\frak a}}_\N}

\def\e{\rm e}
\def\ps{\psi_\N(x)}
\def\bps{\overline{\psi}_\N(x)}
\def\R{R_\N}
\def\I{I_\N}
\def\K{K}
\documentclass[12pt,a4paper]{article}   
\usepackage{graphicx}
\usepackage{amsmath,amsfonts,amssymb}

\title{The spin-1/2 Heisenberg chain: thermodynamics, 
quantum criticality and spin-Peierls
exponents
}
\author{A. Kl\"umper\thanks{e-mail: kluemper@thp.uni-koeln.de}\\
        \parbox{0.9\textwidth}{
        {\em
        \begin{center}
        Universit\"at zu K\"oln  \\                
        Institut f\"ur Theoretische Physik \\     
        Z\"ulpicher Str. 77  \\                   
        D-50937, Germany
        \end{center}
        }}  
       }    

\date{\today}
\begin{document}

\maketitle
\begin{abstract}
We present numerical and analytical results for the thermodynamical properties 
of the spin-1/2 Heisenberg chain at arbitrary external magnetic field. 
Special emphasis is placed on logarithmic 
corrections in the susceptibility and specific heat at very low temperatures
($T/J=10^{-24}$) and small fields. 
A longstanding controversy about the specific heat is resolved. At zero
temperature the spin-Peierls exponent is calculated in
dependence on the external magnetic field. This describes the 
energy response of the system to commensurate and 
incommensurate modulations of the lattice. The exponent for the 
spin gap in the incommensurate phase is given.
\end{abstract}

\noindent
PACS : 75.10.J, 75.40, 72.15.N

\noindent
Short title: ``Heisenberg chain: thermodynamics,
quantum criticality and instabilities''

\clearpage
\section{Introduction}
\label{sec:intro}
The seminal model for correlated quantum many-body systems is the 
well-known Heisenberg model. In one spatial dimension the system 
with nearest neighbour exchange is exactly solvable for the spin-1/2 
case \refto{Bethe31}. The elementary excitations have been 
classified as spinons a long time ago \refto{JohnKM73,FadTak81}.
Despite the integrability 
a comprehensive theoretical treatment is still missing. A notorious
problem is posed by the calculation of correlation functions even at
zero temperature, not to mention the corresponding properties at
finite temperature. However, the asymptotics of the correlations 
in particular in the groundstate and at low temperatures are quite
well understood by a combination of Bethe ansatz \refto{Baxt82b}
and conformal field theory \refto{BelaPZ84,Card88}. 
At exactly zero temperature the spin-1/2 Heisenberg chain
shows algebraic decay in its correlation functions and thereby 
constitutes a quantum critical system.

In recent years several experimental systems have been synthesized
which realize very closely quasi one-dimensional isotropic Heisenberg
chains \refto{Takigawa96,Boeskeetal97}. In experimental studies the two-spinon
continuum characteristic for the one-dimensional geometry
has been verified in the dynamical structure factor,
see for instance \refto{Arai96}. 
Some other 
properties probing the critical properties of the system comprise 
corrections to the low-temperature asymptotics in for instance the 
susceptibility \refto{Takagi96}. 
Typically these corrections are logarithmic singularities
\refto{EggertAT94,OkaNom92,NomOka93,Nom93}
even without impurities. The origin of this behaviour can be traced
back to the existence of marginal operators (in terms of statistical 
mechanics). 

Another aspect of the quantum criticality of the spin-1/2 chain is
its instability towards a structural transition driven by 
long-ranged quantum fluctuations. Such instabilities 
generally known as Peierls or spin-Peierls transitions were 
originally treated in \refto{Peierls} for the case of free fermions
(showing only marginal instability). In these days the understanding
of the truly interacting case has increased in qualitative as well
as quantitative aspects \refto{CrossFisher79,Cross79}. 

In Section 2 of this paper an approach to the thermodynamics
on the basis of the physical excitations (spinons) is 
applied to the study of several thermodynamical properties
\refto{KlumTH} (see also the related approach by
\refto{Koma,Tak91,Bariev82,TruS83,SuzukiAW90,KlumB90,KlumBP91,DestdeVeW92}).
The main results concern the numerical and analytical calculation of the
universal leading and next-leading corrections in the susceptibility.
In particular the
resolution of a controversy between exact \refto{WoynE87}, 
field theoretical \refto{AfflGSZ89} and
numerical approaches \refto{OkaNom92,NomOka93,Nom93,Karbach}
about the specific heat at low temperatures
is presented which is absolutely free of fitting procedures in
the analysis of the zero-temperature limit.
Our approach permits the numerical study of extremely low-temperatures 
down to $T/J=10^{-24}$. Furthermore the analytic treatment
reveals the origin of the logarithmic corrections in certain properties 
of the spinon-spinon scattering phase. In Section 3 we calculate the
critical exponents of correlation functions at zero temperature
in the presence of an external
magnetic field. On the basis of scaling relations the spin-Peierls 
exponents describing the gain in magnetic energy and the opening
of gaps (even in the incommensurate phase) are derived. This is expected
to be the basis of a more accurate treatment of the temperature-field phase
diagram of spin-Peierls systems.

\section{Thermodynamics and logarithmic corrections}

\begin{figure}[tb]
  \begin{center}
    \leavevmode
    \includegraphics[width=0.4\textwidth]{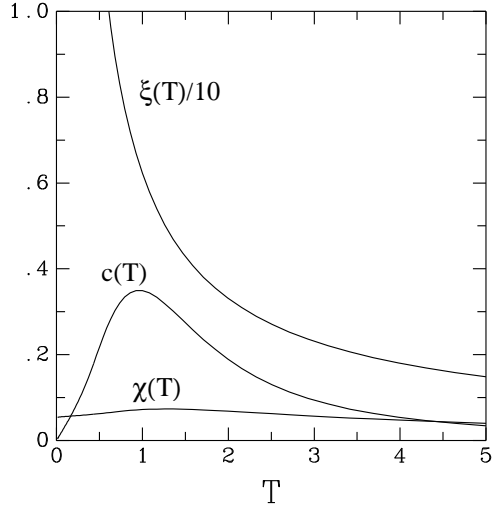}
    \caption{Plots of the specific heat $c(T)$, susceptibility $\chi(T)$
(in units of $1/J$)
and correlation length $\xi(T)$ at zero external field versus
$T$ (in units of $J$).}
    \label{fig:thprop}
  \end{center}
\end{figure}

\begin{figure}[tb]
  \begin{center}
    \leavevmode
    \includegraphics[width=0.3\textwidth]{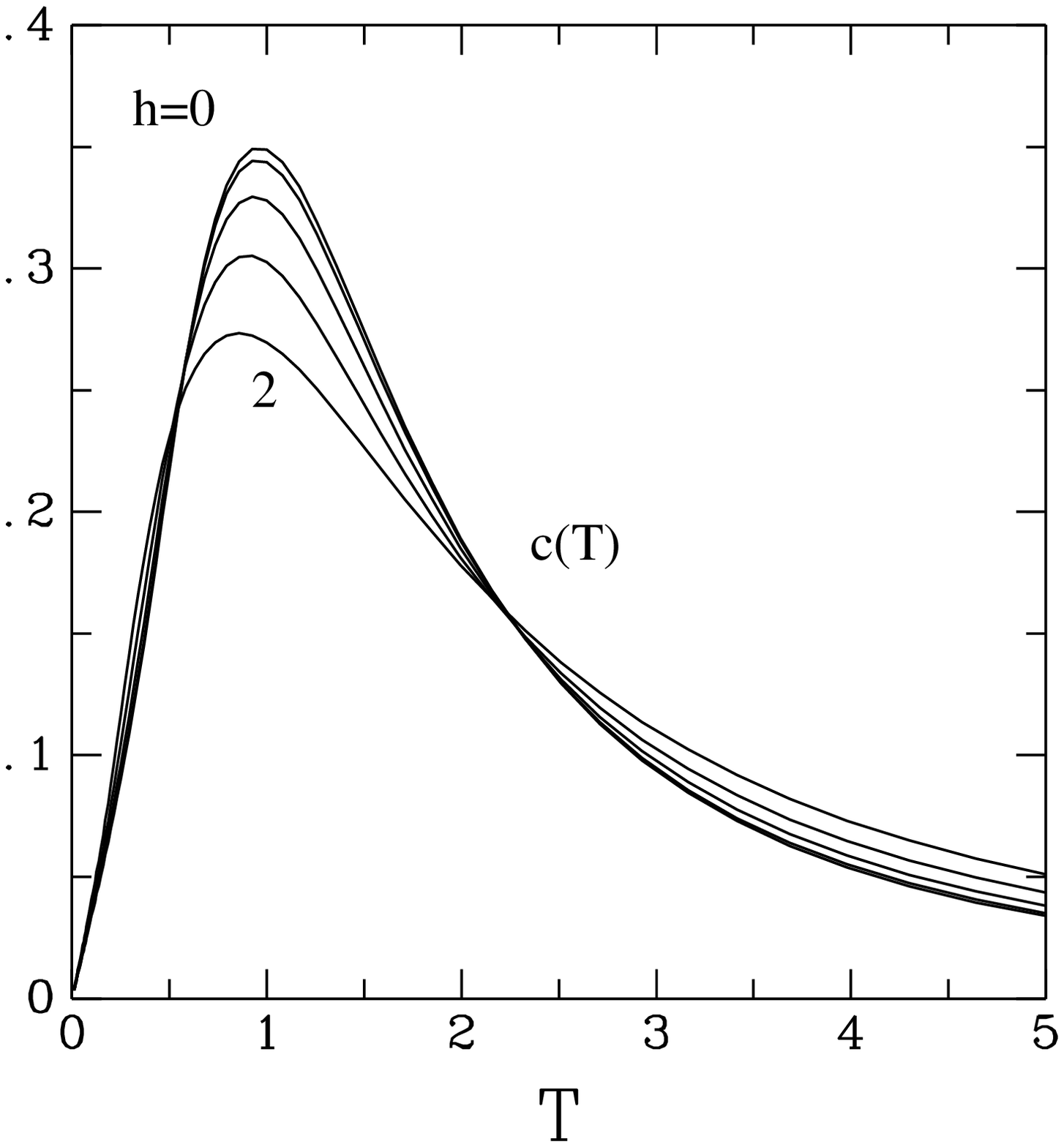}
    \includegraphics[width=0.3\textwidth]{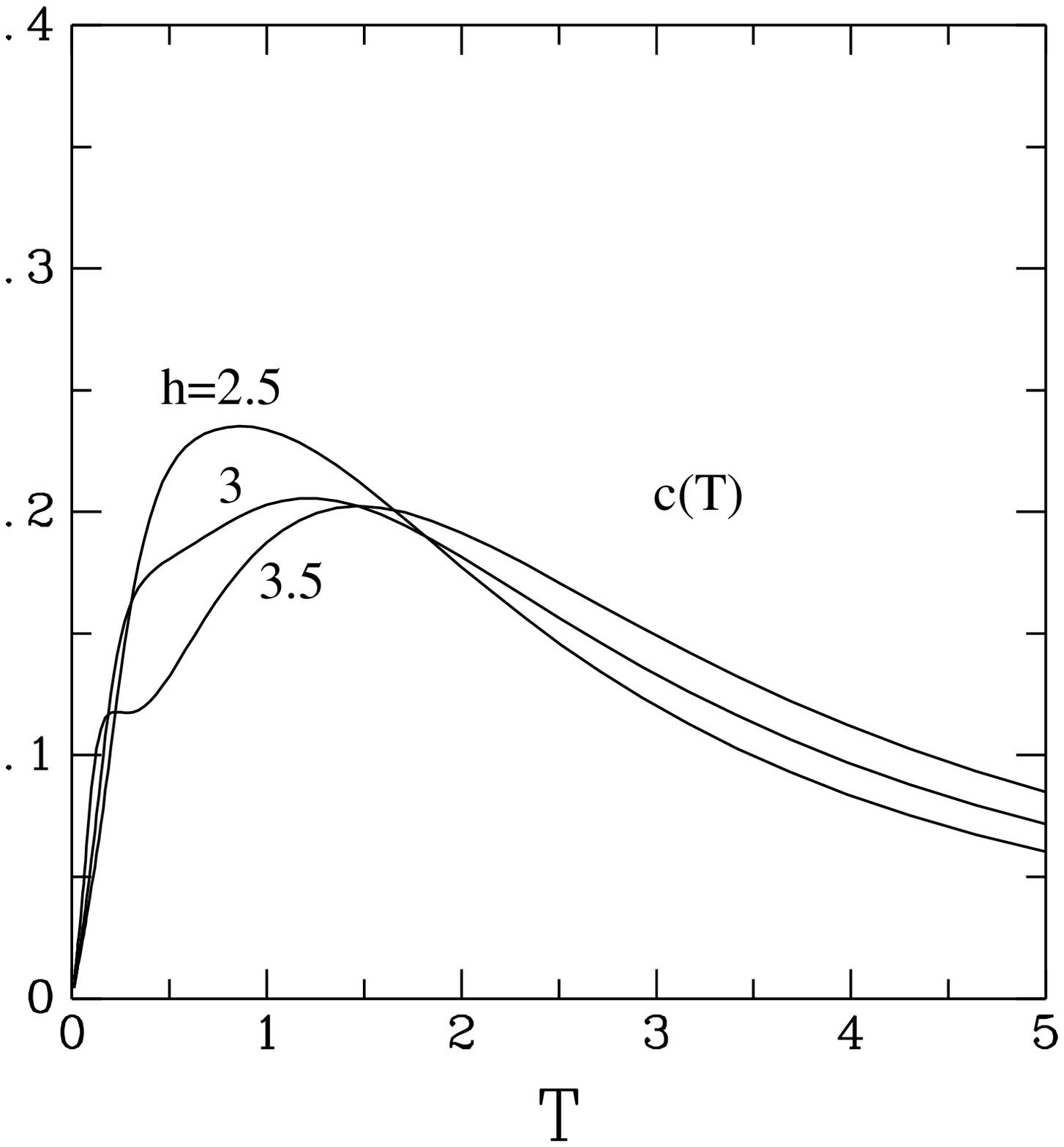}
    \includegraphics[width=0.3\textwidth]{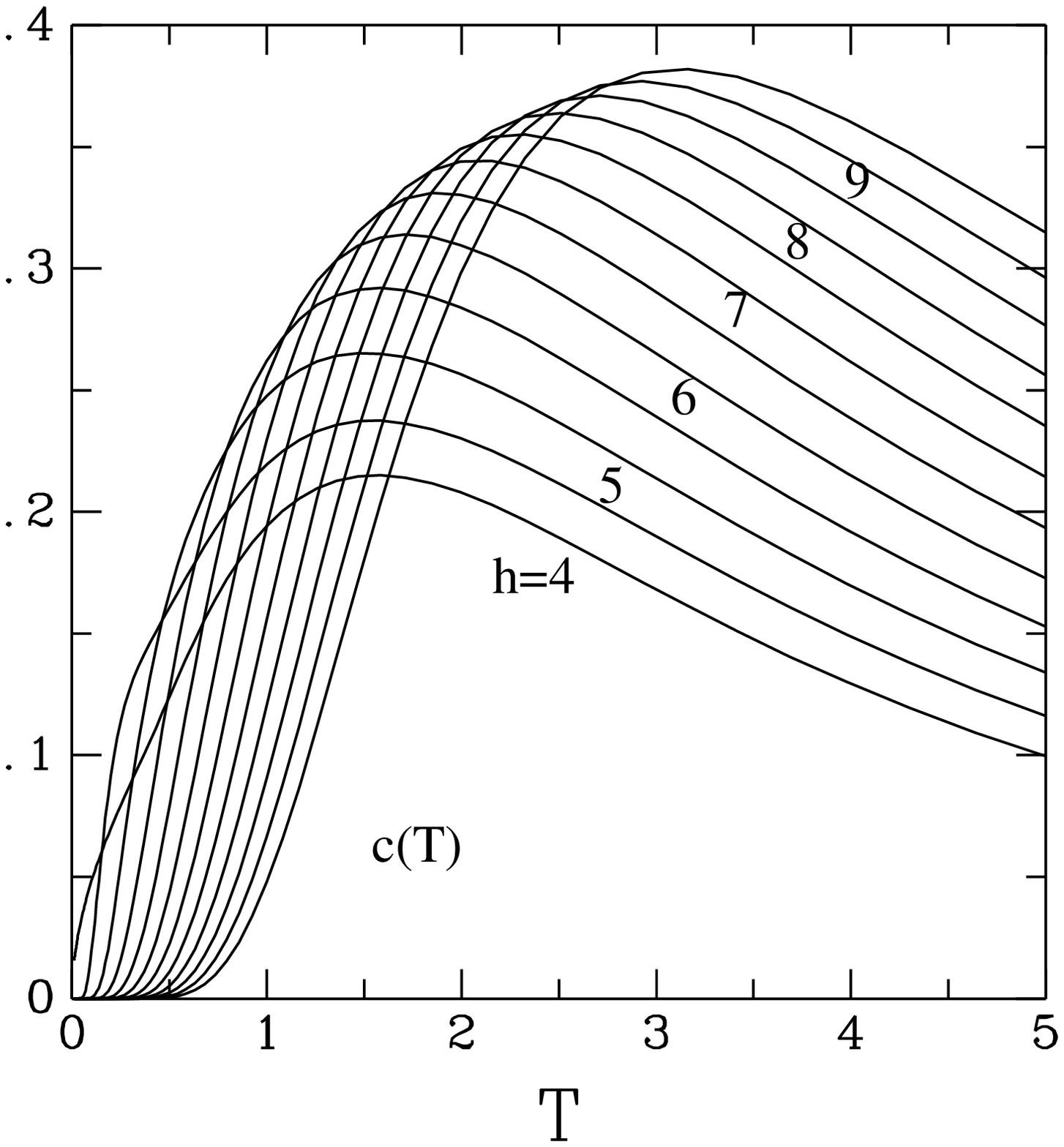}
    \caption{Depiction of the specific heat 
of the Heisenberg chain
for different values of the magnetic field (a) $h/J=0,0.5,...,2.0$,
(b) $h/J=2.5,...,3.5$, (c) $h/J=4.0,...,9.5$. Note the linear $T$
dependence at low temperatures $T$ for fields less than the saturation
value $h_c/J=4$, see (a) and (b). For $h=h_c$ the
specific heat is proportional to $T^{1/2}$ for sufficiently low
$T$, and for even stronger fields
a thermodynamically activated behaviour develops, see (c).}
    \label{fig:spec}
  \end{center}
\end{figure}

\begin{figure}[tb]
  \begin{center}
    \leavevmode
    \includegraphics[width=0.3\textwidth]{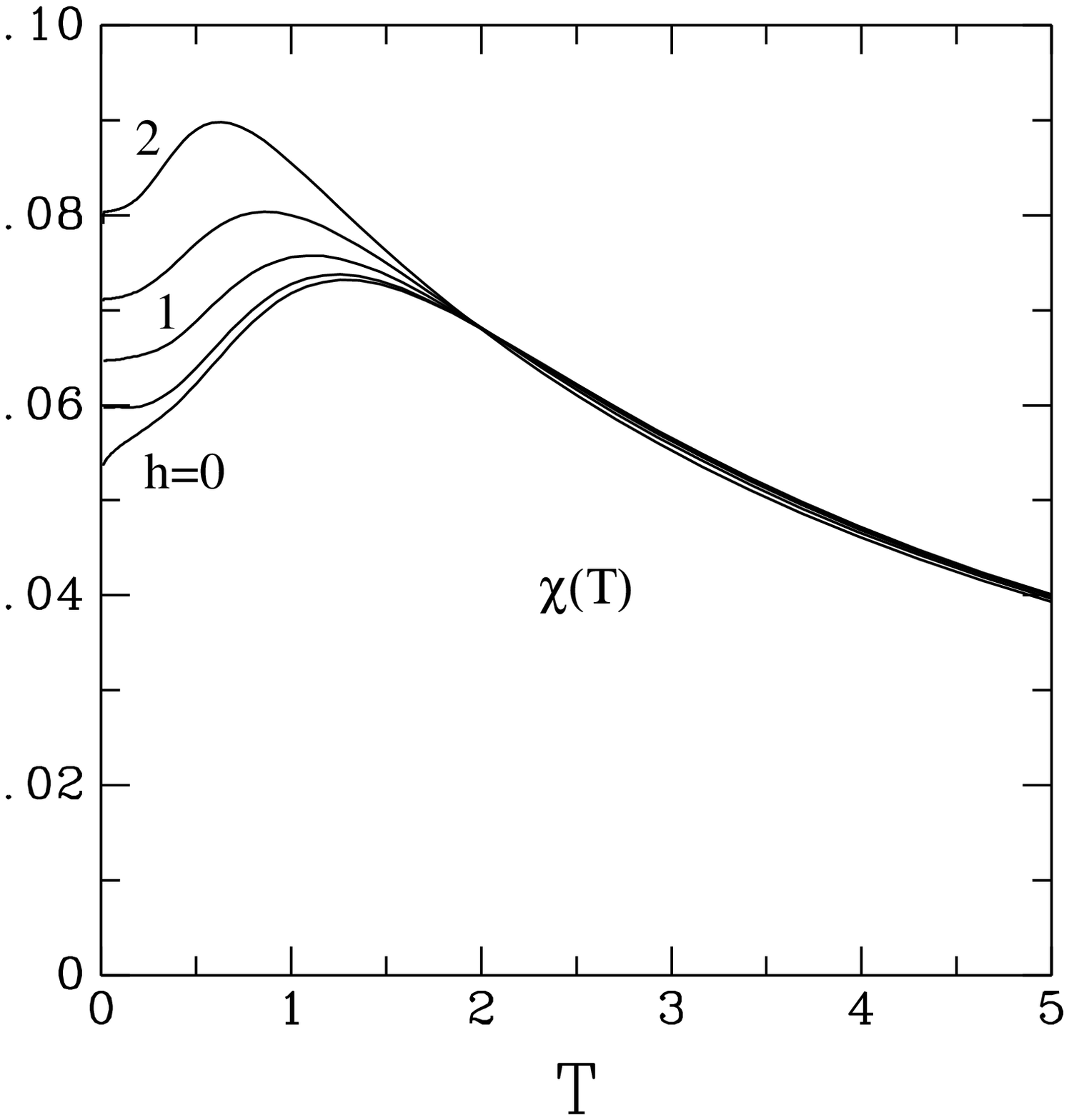}
    \includegraphics[width=0.3\textwidth]{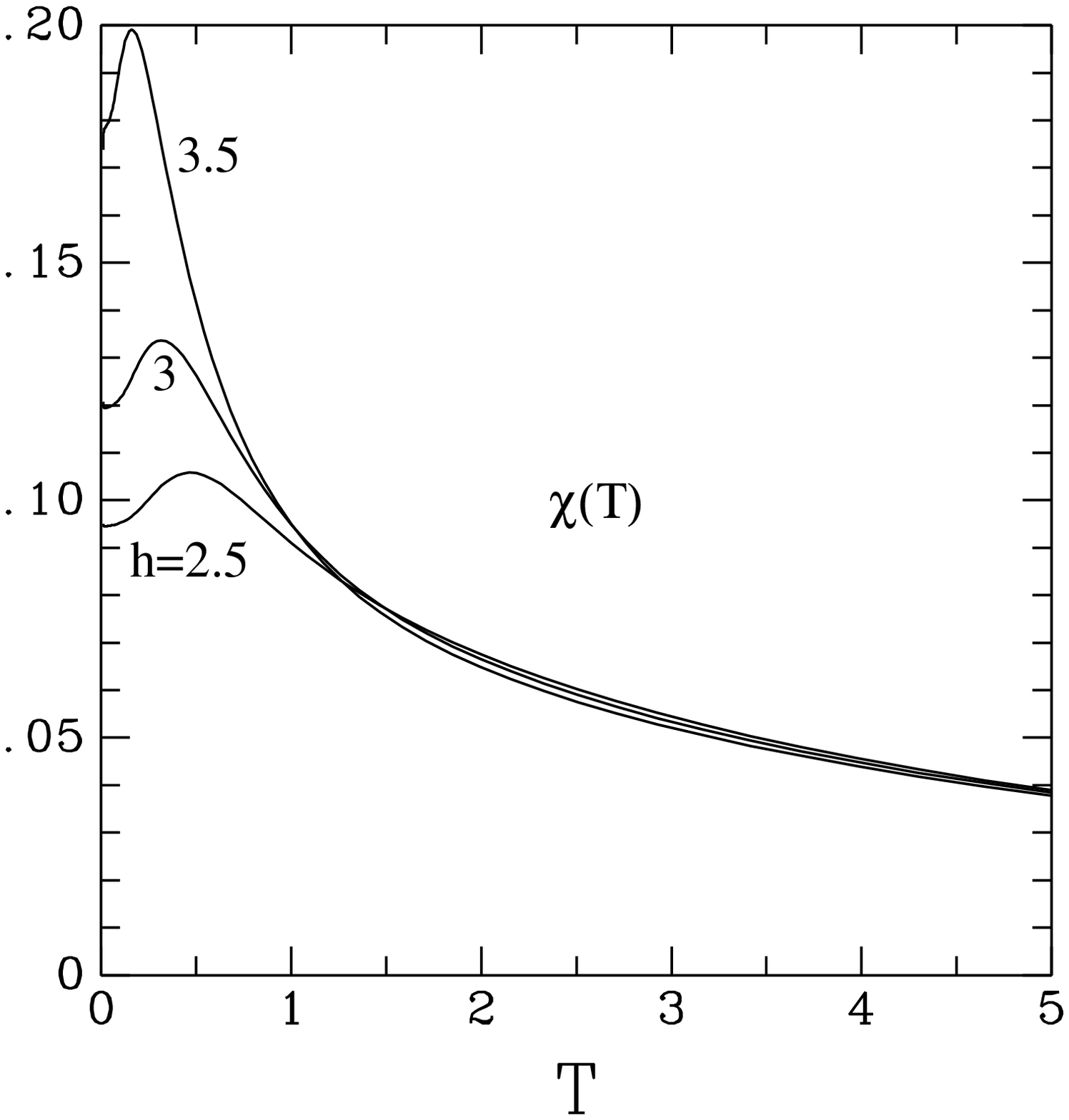}
    \includegraphics[width=0.3\textwidth]{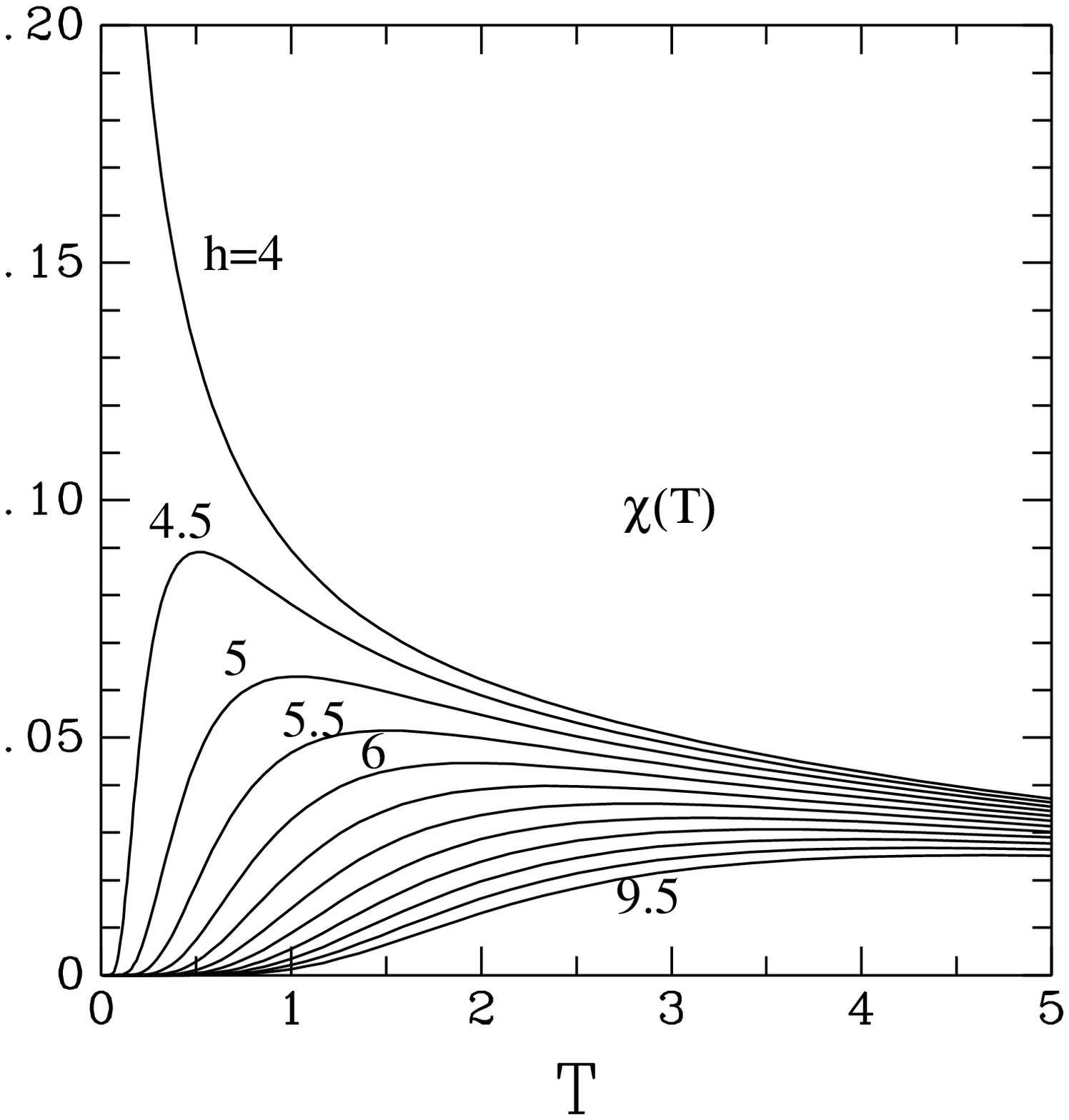}
    \caption{Depiction of the susceptibility (in units of $1/J$)
of the Heisenberg chain
for different values of the magnetic field (a) $h/J=0,0.5,...,2.0$,
(b) $h/J=2.5,...,3.5$, (c) $h/J=4.0,...,9.5$. Note the finite value at
zero temperature for fields less than the saturation
value $h_c/J=4$, see (a) and (b). For $h=h_c$ the susceptibility
diverges at $T=0$, and for fields larger than $h_c$ the susceptibility
drops to zero, see (c).}
    \label{fig:susc}
  \end{center}
\end{figure}

We investigate the thermodynamical properties of the antiferromagnetic
isotropic spin-1/2 Heisenberg chain 
\begin{equation}
H = 2J\sum_{j=1}^L {\vec S}_j{\vec S}_{j+1}.
\end{equation}
We note that the elementary excitations (``spinons'') have quasi-linear
energy-momentum dispersion with velocity
\begin{equation}
v=\pi J.
\end{equation}
The free energy per lattice site of the system is given by the following
set of non-linear integral equations \refto{KlumTH} for auxiliary functions
$\a$, and $\A=1+\a$
\begin{equation}
\log \frak a(x) 
= -\frac{v\beta}{\cosh x}
+\phi+\int_{-\infty}^\infty\left[k(x-y)\log \frak A(y)
-k(x-y-i\pi+i\epsilon)\log\overline {\frak A}(y)\right]{\rm d}y,
\label{IntEqxxx}
\end{equation}
and the corresponding equation for $\ba$, and $\bA=1+\ba$ is obtained from
(\ref{IntEqxxx}) by exchanging 
$\a \leftrightarrow\ba$, 
$\A \leftrightarrow\bA$, and $i \rightarrow -i$,
$\phi\rightarrow -\phi$.  Finally, $\phi$ is given by the external
magnetic field $h$ through $\phi=\beta h/2$.
The integration kernel is defined by the Fourier integral
\begin{equation}
k(x)=\frac{1}{2\pi}\int_{0}^{\infty}
\frac{\e^{-\frac{\pi}{2}k}}{\cosh\frac{\pi}{2}k}\cos(kx)dk.
\end{equation}
In terms of the solution $\A$ and $\bA$ to the integral equations the
free energy is given by
\begin{equation}
\beta f=\beta e_0-\frac{1}{ 2\pi}
\int_{-\infty}^\infty\frac{\log[\A\bA(x)]}{\cosh x}dx.
\label{eigenxxx}
\end{equation}
These equations are readily solved numerically
for arbitrary fields and temperatures
by utilizing the difference type of the integral
kernel and Fast Fourier Transform.
In Fig.~\ref{fig:thprop} the specific heat $c(T)$, 
susceptibility $\chi(T)$ and correlation length $\xi(T)$ (computed on the
basis of \refto{KlumTH}) at zero field are presented. Note the divergence
of $\xi(T)$ at low temperatures.
In Figs.~\ref{fig:spec} and \ref{fig:susc} the results for the specific
heat and susceptibility are shown for various magnetic fields below as well
as above the
saturation field $h_c=4 J$. Some qualitative aspects of the curves
in dependence on the fixed external field can be understood in the 
picture of spinon excitations. The external magnetic field acts very
much like a chemical potential for the spinons for which there are
particle and
hole like excitations. At zero field the bands of the particle and the hole
type excitations are identical, however for sufficiently strong field 
$h$ the band widths are considerably different resulting into two 
maxima in the specific heat at different temperatures.

\begin{figure}[tb]
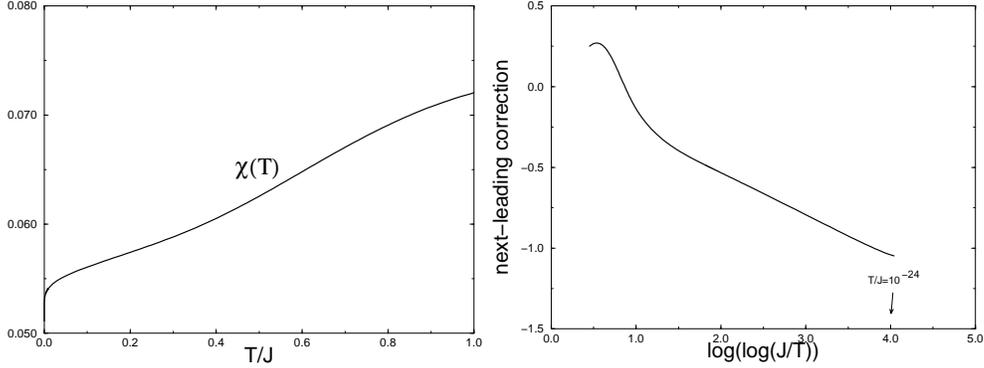

  \begin{center}
    \leavevmode
    \includegraphics[width=0.3\textwidth,angle=270]{Susz.epsi}
    \includegraphics[width=0.3\textwidth,angle=270]{DoubleLog.epsi}
    \caption{Plots of (a) zero-field susceptibility $\chi(T)$ versus $T/J$
showing a logarithmic singularity at $T=0$
and (b) the next-leading logarithmic correction
$(\chi/\chi_0-1-1/2\N)\N^2$ versus $\log\N$ $(\simeq\log\log J/T)$
down to $T/J=10^{-24}$. Note the linear dependence on $\log\N$ for low
temperatures.}
    \label{fig:lowTsusc}
  \end{center}
\end{figure}

In Fig.~\ref{fig:lowTsusc} (a) the susceptibility $\chi(T)$ is plotted down to
much lower temperatures showing the onset of an infinite slope at $T=0$
as noticed in \refto{EggertAT94}. 
In Fig.~\ref{fig:lowTsusc} (b)
a low temperature analysis of $\chi(T)$ down to $T/J=10^{-24}$
is presented giving numerical
evidence about leading and subleading logarithmic corrections in 
temperature. Here we like to present a completely analytical low-temperature
analysis (for $h<<T$) 
and defer comments about the numerical analysis to later.

\subsection*{Analytical study of low-temperature asymptotics}

Because of the driving term on the RHS of (\ref{IntEqxxx}) it is convenient
to introduce the following scaling functions
\bea
\b(x)&=&\a(x+\N),\qquad \N=\log(\pi\beta),\nonumber\\
\tb(x)&=&\a(-x-\N),
\eea
approaching well-defined non-trivial limiting functions in the low-temperature
limit which satisfy
\bea
\log \b(x) 
&=& -2\e^{-x}+O\left({1}/{\beta}\right)
+\phi+\ps\nonumber\\
&&+\int_{-\N}^\infty\left[k(x-y)\log \B(y)
-k(x-y-i\pi+i\epsilon)\log\bB(y)\right]{\rm d}y, 
\label{IntEqxxxl}
\eea
where $\ps$ accounts for the contribution of the functions $\A$ and $\bA$
on the negative real axis
\begin{equation}
\ps=\int_{-\N}^\infty\left[k(x+y+2\N)\log \tB(y)
-k(x+y+2\N-i\pi+i\epsilon)\log\tbB(y)\right]{\rm d}y.
\label{ps}
\end{equation}
Using these functions the free energy at low temperatures
takes the form
\begin{equation}
f=e_0-\frac{1}{\pi v\beta^2}
\int_{-\N}^\infty\e^{-x}\log\left[\B\bB\tB\tbB(x)\right]dx
+O\left(1/\beta^3\right).
\label{finiteT}
\end{equation}
In order to analyse the low-temperature asymptotics in detail we
perform the following manipulation 
\bea
&&\int_{-\N}^\infty\left[(\log\b)'(\log\B)-(\log\b)(\log\B)'\right]dx
+\hbox{`conj'},
\nonumber\\
&&=\int_{-\N}^\infty\left[(-2\e^{-x}+\phi+\ps)'(\log\B)-
(-2\e^{-x}+\phi+\ps)(\log\B)'\right]dx+\hbox{`conj'},\nonumber\\
&&=2\int_{-\N}^\infty(2\e^{-x}+\ps')(\log\B)dx-
\phi\log\left(1+\e^{2\phi}\right)+\hbox{`conj'},
\label{eq1}
\eea
where in the second line we have inserted (\ref{IntEqxxxl}) and in the
third line we have performed an integration by parts with an explicit
evaluation of the contribution
by the integral terminals using the limits
\bea
\log\b(\infty)&=&{+2\phi},\qquad \log\B(\infty)=
\log\left(1+\e^{+2\phi}\right),\nonumber\\
\log\bb(\infty)&=&{-2\phi},\qquad  
\log\bB(\infty)=\log\left(1+\e^{-2\phi}\right).
\label{limits}
\eea
Next, we show that the LHS of (\ref{eq1}) can be evaluated explicitly
as it is a definite integral of dilogarithmic type with known terminals
\bea
LHS&=&2\int_{-\N}^\infty(\log\b)'\log\B dx-
\log\b(\log\B)'\Big|_{-\N}^\infty+\hbox{`conj'},\nonumber\\
&=&2\int_{-\infty}^{2\phi}\log(1+\e^z)dz-2\phi\log(1+\e^{2\phi})
+2\int_{-\infty}^{-2\phi}\log(1+\e^z)dz+2\phi\log(1+\e^{-2\phi})
,\nonumber\\
&=&\frac{\pi^2}{3},
\label{eq2}
\eea
and the final result for the definite integral in the second line 
was obtained by standard techniques
(it is independent of $\phi$ and the
particular choice $\phi=0$ ``directly'' leads to the final result).

Combining (\ref{eq1}) and (\ref{eq2}) we arrive at a formula
particularly suited for studying the asymptotic behaviour of
(\ref{finiteT})
\begin{equation}
4\int_{-\N}^\infty\e^{-x}\log[\B\bB]dx=
\frac{\pi^2}{3}+2\phi^2-2\int_{-\N}^\infty\left[\ps'\log\B+\bps'\log\bB
\right]dx.
\label{eq3}
\end{equation}
From the general expression of the free energy $f$ (\ref{eigenxxx})
in terms of the functions $\A$ and $\bA$ the entire functional
dependence of $\A$ and $\bA$ matters even for the calculation of
the asymptotic behaviour of $f$ at low temperature.
The usefulness of (\ref{eq3}) lies in the fact that here only the 
asymptotic behaviour of $\A$ and $\bA$ at large spectral parameter
enters. The reason is simple to understand: for any finite argument $x$
the contribution is of order $1/\N^2$ due to the $\ps$ factors.
Here we see that the low-temperature
asymptotics (and further below the occurrence of logarithmic
corrections) are intimately linked to the algebraic decay of the
integration kernel $k(x)$
\begin{equation}
k(x)\simeq\frac{1}{\pi^2+4x^2}.
\end{equation}
We like to add that the kernel is related to the $S$-matrix 
of the elementary spinons \refto{FadTak81}
by $k(x)=\frac{d}{dx}\log S(x)$.

\subsection*{Asymptotics of the susceptibility}

In a first step we determine the asymptotics of the
auxiliary functions from (\ref{IntEqxxxl}) where 
we drop integrations over negative arguments
and for large arguments we use
$\bb\simeq 1/\b$ (hence $\log\B-\log\bB=\log\b$) 
and furthermore we 
equate the kernels $k(...)$ with arguments differing by $\pi i$
leading to
\bea
\log \b(x) 
&=& \phi+\ps+\int_{0}^\infty k(x-y)\log \b(y){\rm d}y, \nonumber\\
&=& \phi+\ps+\int_{0}^\infty k(x-y)\log \frac{\b(y)}{\b(\infty)}{\rm d}y+
\int^{\infty}_0 k(x-y)\log \b(\infty){\rm d}y, \nonumber\\
&=& \phi+\ps+\log \frac{\b(x)}{\b(\infty)}\int_{0}^\infty k(x-y){\rm
d}y
\nonumber\\
&&+\int_{-\infty}^\infty k(x-y)\log \b(\infty){\rm d}y-
\int_{-\infty}^0 k(x-y)\log \b(\infty){\rm d}y, \nonumber\\
&=& \phi+\ps+\frac{1}{2}\log \b(x)-
\int_{-\infty}^0 \frac{\log \b(y)}{4(x-y)^2}{\rm d}y, \nonumber\\
&=& \phi+\ps+\frac{1}{2}\log \b(x)-\frac{\phi}{2x}.
\eea
In the first integral of the second line we have used the fact that
dominant contributions occur for the integration variable $y$ close to $x$.
In the subsequent calculations the asymptotic behaviour of the kernel
and also the total integral $\int k(x)dx=1/2$
were used. Finally, employing the limits
(\ref{limits}) we arrive at
\begin{equation}
\log \b(x)=2\phi-\frac{\phi}{x}.
\label{asy}
\end{equation}

In the second step we insert (\ref{ps}) into the integral on
the RHS of (\ref{eq3}), then we insert (\ref{asy}) and calculate
the resulting integral by dropping terms leading to contributions
of order $O(1/\N^2)$ and higher
\bea
&&\int_{-\N}^\infty\left[\ps'\log\B+\bps'\log\bB
\right]dx\nonumber\\
&&=+4\phi^2\int_{1}^\infty\int_{1}^\infty 
k'(x+y+2\N)\left(1-\frac{1}{2x}\right)\left(1-\frac{1}{2y}\right)
dxdy\nonumber\\
&&=-4\phi^2\int_{1}^\infty  k(x+2\N)\left(1-\frac{1}{x}\right)dx
=-\phi^2\int_{1}^\infty \frac{1}{(x+2\N)^2}\left(1-\frac{1}{x}\right)
dx\nonumber\\
&&=\phi^2\left(-\frac{1}{2\N}+\frac{\log(2\N)}{(2\N)^2}\right).
\eea
Hence we obtain the dependence of the free energy on the magnetic field
at low temperature
\begin{equation}
4\int_{-\N}^\infty\e^{-x}\log[\B\bB]dx=
\frac{\pi^2}{3}+2\phi^2\left(1+\frac{1}{2\N}-\frac{\log(2\N)}{(2\N)^2}\right),
\end{equation}
implying logarithmic corrections in $T$ for the free energy and the 
zero-field susceptibility
\bea
f&=&e_0-\frac{\pi}{6v}T^2-
\frac{h^2}{4\pi v}\left(1+\frac{1}{2\N}-\frac{\log(2\N)}{(2\N)^2}\right),
\quad\N=\log(\pi/T),\quad (h<<T),\\
\chi&=&\frac{1}{2\pi v}\left(1+\frac{1}{2\N}-\frac{\log(2\N)}{(2\N)^2}\right).
\eea
This result is in complete agreement with the numerical analysis in
Fig.~\ref{fig:lowTsusc} (b) and with \refto{KawanoT95,Karbach} (if the
system size $L$ is replaced by the inverse temperature $1/T$). 
The (non-universal) temperature scale $T_0$
entering the logarithms of the correction terms will be studied elsewhere
\refto{KlumTak98}.

\subsection*{Asymptotics of the specific heat}

\begin{figure}[tb]
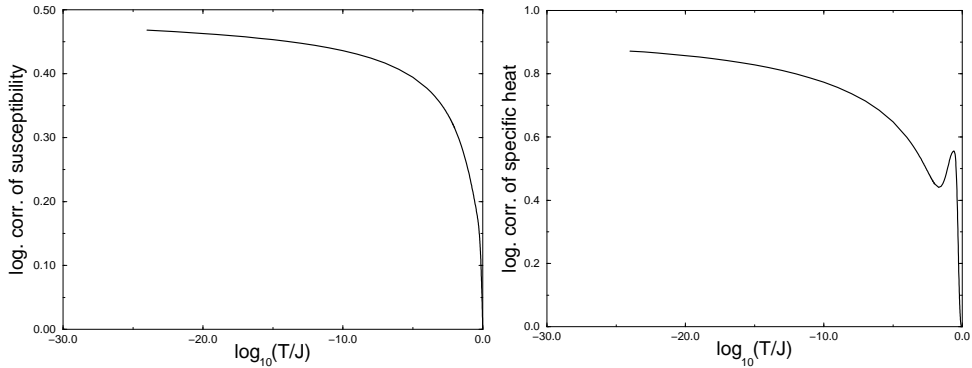

  \begin{center}
    \leavevmode
    \includegraphics[width=0.3\textwidth,angle=270]{susclog.epsi}
    \includegraphics[width=0.3\textwidth,angle=270]{spezlog.epsi}
    \caption{Plots of the estimators of the leading logarithmic
corrections (a) for the susceptibility $(\chi/\chi_0-1)\N$ and 
(b) the specific heat $[3v c(T)/\pi T-1]\frac{8}{3}\N^3$ versus 
$\log_{10}(T/J)$.
Note the non-monotonous behaviour of the higher order terms in (b).}
    \label{fig:estimators}
  \end{center}
\end{figure}

Next we calculate the higher order 
low-temperature asymptotics of the Heisenberg
chain in zero external magnetic field. Here the functions $\b$ and $\bb$
are related by complex conjugation. We therefore write
\bea
\log\B(x)&=&\log\tbB(x)=\R(x)+i\I(x),\nonumber\\
\log\bB(x)&=&\log\tB(x)=\R(x)-i\I(x).
\label{cc}
\eea
Again we insert (\ref{ps}) into the integral on
the RHS of (\ref{eq3}), then we insert (\ref{cc}) and 
expand $k(x+\pi i)=k(x)+k'(x)\pi i-k''(x)\pi^2/2$ obtaining
\bea
&&\int_{-\N}^\infty\left[\ps'\log\B+\bps'\log\bB\right]dx\nonumber\\
&&=\int_{-\N}^\infty\int_{-\N}^\infty dx dy\big[
\pi^2 k'''(x+y+2\N)\R(x)\R(y)+4 k'(x+y+2\N)\I(x)\I(y)
\nonumber\\
&&\phantom{=\int_{-\N}^\infty\int_{-\N}^\infty dx dy\big[}
-2\pi k''(x+y+2\N)\left(\R(x)\I(y)+\I(x)\R(y)\right)\big],\nonumber\\
&&=\int_{-\N}^\infty\int_{-\N}^\infty dx dy\big[
\pi^2 k'(x+y+2\N)\R'(x)\R'(y)+4 k'(x+y+2\N)\I(x)\I(y)
\nonumber\\
&&\phantom{=\int_{-\N}^\infty\int_{-\N}^\infty dx dy\big[}
+2\pi k'(x+y+2\N)\left(\R'(x)\I(y)+\I(x)\R'(y)\right)\big],
\eea
where in the last step we have performed an integration by parts. Next,
the integral is dominated by contributions with finite integration
variable $x$ and $y$ for which $k'(...)$ can be replaced by a constant
yielding
\bea
RHS&=&-\frac{1}{16\N^3}\int_{-\N}^\infty\int_{-\N}^\infty dx dy\big[
\pi\R'(x)+2 \I(x)\big]\big[\pi \R'(y)+2\I(y)\big],
\nonumber\\
&=&-\frac{1}{16\N^3}(\pi R+2 I)^2,
\eea
where we have used the abbreviations
\bea
R&=&\int_{-\N}^\infty dx\R'(x)=\log 2,\nonumber\\
I&=&\int_{-\N}^\infty dx\I(x)\to \hbox{ numerical evaluation},
\eea
where $R$ can be calculated easily from the asymptotics of $\R(x)$,
$I$ however has to be evaluated by numerical calculations. For the free
energy we obtain
\begin{equation}
4\int_{-\N}^\infty\e^{-x}\log[\B\bB]dx=
\frac{\pi^2}{3}\left[1+\frac{3}{8}
\left(R+\frac{2}{\pi}I\right)^2\frac{1}{\N^3}\right],
\label{logspech}
\end{equation}
where $\left(R+\frac{2}{\pi}I\right)$ is found to be $1\pm 10^{-6}$.
This leads to
\begin{equation}
f=e_0-\frac{\pi}{6v}T^2\left(1+\frac{3}{8}\frac{1}{\N^3}\right),
\quad\N=\log(\pi/T),
\end{equation}
confirming the field theoretical
prediction in \refto{AfflGSZ89} about the amplitude of the $1/(\log T)^3$
correction which was argued to be universal however in disagreement
with Bethe ansatz calculations in \refto{WoynE87}. The reason for the
``failure'' of the treatment in \refto{WoynE87} (instead of $3/8$ the value
$0.3433$ was obtained) is subtle and lies in an inappropriate use of the
Euler-MacLaurin formula which unfortunately is common practice in the
treatment of critical lattice systems. The numerical treatments of
the Bethe ansatz equations up to $N=16384$ \refto{OkaNom92,NomOka93,Nom93}
were still plagued by 
higher order logarithmic contributions leading to erroneous conclusions
with respect to the analysis of the specific heat. For an illustration
of these higher order terms see Fig. \ref{fig:estimators}. For the 
case of the specific heat an appropriate fit of finite size data seems
very difficult. We note, that the careful numerical treatment \refto{Karbach}
of finite systems up $N=16384$ employed an appropriate fit algorithm
taking into account the proper higher correction terms. Our results agree
with the conclusions of \refto{Karbach}.
We emphasize that our treatment is completely free of any fit procedures.
The type of logarithmic singularities has been derived analytically and
the amplitudes have been calculated (in the worst case) numerically
which did not face any extrapolation problems.

\section{Groundstate properties of homogeneous and dimerized S=1/2 
Heisenberg chains}
Next we treat the critical groundstate properties of the homogeneous 
Heisenberg chain by means of an application of the exact Bethe ansatz solution
and conformal field theory, see
\refto{KorBI93} and references therein. 

\subsection*{Bulk properties}

Directly from the 
Bethe ansatz equations
for spin rapidities the (linear) integral equation for the groundstate 
density function $\rho(v)$ is derived \refto{Baxt82b,KorBI93}
\begin{equation}
\rho(x)=\rho_0(x)-\int_{-\K}^\K k(x-y)\rho(y)dy,
\label{lintegraleq}
\end{equation}
where the so-called bare density $\rho_0$ and the kernel $k(x)$ are defined
by
\begin{equation}
\rho_0=\frac{1}{2\pi}\frac{2}{x^2+1},\quad
k(x)=\frac{1}{2\pi}\frac{4}{x^2+4}.
\end{equation}
Here we have as yet left unspecified the magnetization $m$ of the state 
(or equivalently the external magnetic field $h$) corresponding 
to some integration terminal $\K$ with subsidiary condition 
\begin{equation}
m=\frac{1}{2}-\int_{-\K}^\K \rho(x)dx,\quad 2 k_F=(1-2m)\pi,
\label{mag}
\end{equation}
where (twice) the Fermi momentum $k_F$ is the momentum at which 
elementary spin excitations become soft. 
In terms of the density function the groundstate energy $e$ takes the form
\begin{equation}
e=\int_{-\K}^\K \epsilon_0(x)\rho(x)dx,
\label{gse}
\end{equation}
where $\epsilon_0(x)$ is the bare energy of spin excitations
\begin{equation}
\epsilon_0(x)=-\frac{4J}{x^2+1},\quad
\xi_0(x)=1,
\label{bare}
\end{equation}
and the bare charge function $\xi_0(x)$ has been introduced for later purposes.

In the same way as (\ref{lintegraleq}) defines a function $\rho$ for 
the inhomogeneity $\rho_0$, it gives rise to the dressed energy $\epsilon(x)$
and the dressed charge $\xi(x)$ related to (\ref{bare}). (Note that 
the auxiliary function $\a(x)$ of the thermodynamic treatment 
is related to $\epsilon(x)$ in the low-temperature limit via $\log\a=
\beta\epsilon$ and (\ref{IntEqxxx}) turns into the linear integral
equation for the dressed energy.) As the integration
operator in (\ref{lintegraleq}) is selfadjoint we have an alternative
relation to (\ref{gse}) and (\ref{mag}) for the groundstate energy $e$
and magnetization $m$
\begin{equation}
e=\int_{-\K}^\K \epsilon(x)\rho_0(x)dx,\qquad
m=\frac{1}{2}-\int_{-\K}^\K \xi(x)\rho_0(x)dx.
\label{emag}
\end{equation}

For relating the terminal $\K$ to the magnetic
field $h$ we note that the bare energy in the presence of
an external field is given by
$\epsilon_0(x)+h=\epsilon_0(x)+ {h} \xi_0(x)$. Hence 
the corresponding dressed energy is
$\epsilon(x,h)=\epsilon(x)+{h} \xi(x)$.
A general property of the dressed
energy is that it has to vanish at $\K$ yielding
\begin{equation}
\epsilon(\K,h)=0 \Leftrightarrow h=-\frac{\epsilon(\K)}{\xi(\K)}.
\label{field}
\end{equation}
For any value of $\K$ the functions $\epsilon(x)$, $\xi(x)$ can be evaluated
(at least numerically) from an integral equation of type (\ref{lintegraleq}), 
then the magnetic field $h$, and the corresponding magnetization $m$
and groundstate energy $e-m h$ can be calculated utilizing (\ref{field})
and (\ref{emag}).

\subsection*{Correlation functions}

We are particularly interested in the determination of scaling dimensions
(critical exponents) for arbitrary external fields. 
For instance the spin-spin and energy-energy correlations decay algebraically 
with exponents $x_s$ and $x_e$
\begin{equation}
\langle{\vec S_0}{\vec S_r}\rangle\simeq\frac{(-1)^r}{r^{2x_s}},\quad
\langle\epsilon_0\cdot\epsilon_r\rangle\simeq
\frac{\cos(2k_Fr)}{r^{2x_e}}+\frac{C}{r^{2}},
\end{equation}
where the energy operator is $\epsilon_r={\vec S_r}{\vec S_{r+1}}$.
In general the exponents are 
directly given in terms of the dressed charge function at the terminal 
$\K$ (see \refto{KorBI93}) 
\begin{equation}
x=\frac{1}{4\xi(\K)^2}S^2+\xi(\K)^2 d^2,\label{scaldim}
\end{equation}
where $S$ and $d$ are integers corresponding to the quantum numbers of
the particular operator (occurring in the correlation function); 
$S$ equals the magnetization and $d$ equals the lattice
momentum in units of $2k_F$. For spin and energy operators (correlations)
we have $(S,d)=(1,0)$ and $(S,d)=(0,1)$, respectively. Therefore we obtain
in terms of the dressed charge
\begin{equation}
x_s=\frac{1}{4\xi(\K)^2},\quad x_e=\xi(\K)^2.
\end{equation}

We want to note that (\ref{scaldim}) is obtained from 
predictions by conformal field theory
about finite-size effects for the spectrum of the system
on large lattices $L$ with periodic boundary conditions
\begin{equation}
E_x-E_0=2\pi\frac{v}{L}(x+\delta(L)),\label{finitesize}
\end{equation}
where $E_0$ is the groundstate energy, $E_x$ the low-lying energy level
corresponding to the operator with dimension $x$, 
$v$ the velocity of the elementary excitations, and $\delta(L)$ is
a function with zero limit for $L\to\infty$. The evaluation of 
(\ref{finitesize}) can be carried out analytically yielding
(\ref{scaldim}) (see for instance 
\refto{KorBI93} and references therein). Of course, the dressed
charge $\xi$ has to be calculated numerically.
Alternatively, similar
results can be obtained by numerical
calculations on the basis of the Bethe ansatz equations for finite systems
and (\ref{finitesize}), however involving an explicit extrapolation
analysis with the usual disadvantages, 
see \refto{AlcaBB88} and for finite external fields 
\refto{FleddGMSK96}.

\subsection*{Perturbations of the pure system}

The algebraic decay of the correlation functions in the 
groundstate of the spin-1/2 Heisenberg chain constitutes 
quantum criticality. Starting from this observation 
we are going to argue that the
response of the system to perturbations can be understood and evaluated
in a renormalization group framework and in general shows
non-integer exponents (in contrast to simple perturbation treatments
which of course are not legitime).
We want to examine the critical exponents of the Spin-Peierls transition
which is a structural transition of the underlying lattice
with a modification of the local exchange couplings proportional to
the local lattice distortion
\begin{equation}
 H(\delta)= { H}+{{\sum_{j}}}\delta_j\cdot{\vec S_j}{\vec S_{j+1}}
={ H}+\delta  H.
\label{pert}
\end{equation}
An application of scaling relations will provide a 
simple tool to understand some of the essential aspects of the
theory of \refto{CrossFisher79,Cross79}. We remind the reader of the response of the 
free energy $f$ and the correlation length $\xi$ of a critical, 
classical system in $d$ dimensions perturbed by a relevant operator
with RG eigenvalue $y (>0)$

\medskip
\hbox to \hsize{
\fbox{\vbox{\hbox{}\hbox{classical system}\hbox{$d$ dimension}\hbox{}}}
\vbox{\hbox{}\hbox{$\longrightarrow$}\hbox{}}
\fbox{\vbox{\hbox{${\cal H}(\delta)={\cal H}+\delta\cdot{\cal H}'$}
\hbox{free energy $\Delta f\sim\delta^{\mbox{$\frac{d}{y}$}}$}
\hbox{corr. length $\xi^{-1}\sim\delta^{\mbox{$\frac{1}{y}$}}$}}}
}
\medskip

A quantum critical system in $d$ dimensions behaves very much the same 
as it is equivalent to a $d+z$ dimensional classical system, where $z$ is
the dynamical critical exponent. 

\medskip
\hbox to \hsize{
\fbox{\vbox{\hbox{quantum critical system}\hbox{$d$ dimension}
\hbox{g.s. energy $e$, gap $m_0$}}}
\vbox{\hbox{}\hbox{$\longleftrightarrow$}\hbox{}}
\fbox{\vbox{\hbox{classical system}
\hbox{$d+z$ dimension}
\hbox{$f$, $\xi^{-1}$}}}
}
\medskip

\begin{figure}[tb]
  \begin{center}
    \leavevmode
    \includegraphics[width=0.6\textwidth]{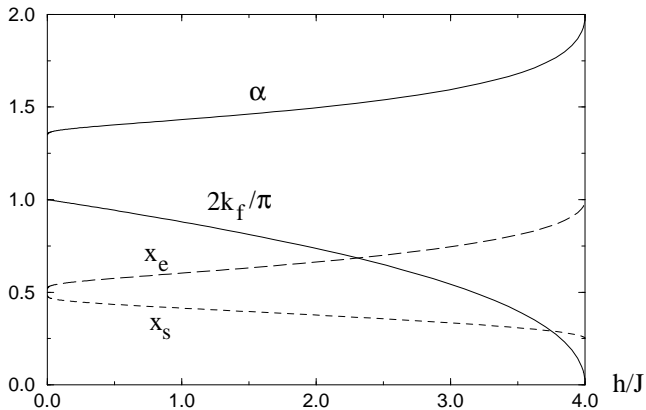}
    \caption{Depiction of scaling dimensions $x_s$, $x_e$, groundstate energy
exponent $\alpha$, and Fermi momentum $k_F$.}
    \label{fig:fig1}
  \end{center}
\end{figure}

\begin{figure}[tb]
  \begin{center}
    \leavevmode
    \includegraphics[width=0.2\textwidth]{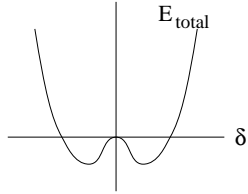}
    \caption{Competition of gain in magnetic energy with loss in elastic 
energy $E_{\rm total}(\delta)=E_{\rm mag}+E_{\rm
elast}=-c_1\delta^\alpha+
c_2\delta^2$.}
    \label{fig:fig2}
  \end{center}
\end{figure}

For conformally invariant chains we
have $d=z=1$. Furthermore the RG eigenvalue $y$ is related to the scaling 
dimension $x$ of the particular operator by 
\begin{equation}
x+y=d+z=2.
\end{equation}
Now equating the above relations we obtain for the response of the 
quantum chain in the ground state energy per site $e$ and the excitation
gap $m_0$ ($=0$ at the unperturbed point):
\begin{equation}
\Delta e\simeq -\delta^{\mbox{$\frac{2}{2-x}$}},\qquad
m_0\simeq \delta^{\mbox{$\frac{1}{2-x}$}}.
\end{equation}
For the spin-Peierls transition we have to use $x=x_e=\xi(\K)^2$. 
In Fig.~\ref{fig:fig1} the value of the groundstate exponent $\alpha=2/(2-x_e)$
is depicted and shows a variation from the value $4/3$ for zero
external field to $2$ at the saturation field. As the exponent is less than
$2$ the gain in magnetic energy always wins in competition with the
loss in elastic energy which is of order $\delta^2$ resulting in
a structural instability, cp. Fig.~\ref{fig:fig2}. Of course this
result is completely consistent with \refto{CrossFisher79,Cross79}
where bosonization results were used for the energy-energy
correlations of the spin-1/2 $XXZ$ chain in vanishing external field
in combination with scaling relations for the polarizability.
In our treatment we have correctly accounted for the field dependence
of the exponents which is new. 

Our study is partly motivated by the
general interest in a deeper understanding of the physics of 
Spin-Peierls systems. 
Here we have shown how a combination of established results by
conformal field theory, RG theory and integrable systems 
yields the exact critical exponents.
At the same time we intend to use our results 
for a more precise calculation of for instance the phase diagram
of Spin-Peierls systems in dependence on temperature and magnetic field.
Note that there are quantitative discrepancies 
in for instance 
the phase boundary between the incommensurate and the uniform
phase as obtained experimentally \refto{KirKei96,Lorenz97}
and theoretically \refto{Cross79}, cf. the remark in 
\refto{Lorenz97}. In addition, we note that the occurence of a
gap in the magnetic system (for fixed modulation of the exchange
couplings) is a very natural consequence of our approach. Experimentally it is
still an open question if a spin gap in the incommensurate phase exists.
The $T^3$ contributions to the specific heat at low temperatures 
\refto{Liu95,Weiden95}
are explained in terms of gapless ``phasons'' \refto{BeNaRon97}.

Up to now we have not specified any particular dependence of the perturbation
field $\delta_j$ on the site index $j$. The above argument 
(based on a relevant
perturbation) is applicable just for the case of a simple algebraic decay
of $\langle (\delta_i\epsilon_i)(\delta_j\epsilon_j)\rangle$ for 
$(i-j)\to\infty$
without any oscillations which imposes a matching condition: 
(i) the modulation is sinusoidal $\delta_j=\delta\cos (q j)$, and
(ii) the wave number $q$ is identical to 
that of the energy-energy correlation 
$q=2k_F$. (For sinusoidal modulations with different wave number the
response in the groundstate energy has the exponent 2.)

As is indicated by the results in Fig.~\ref{fig:fig1} 
the external magnetic field 
leads to a filling of the spinon-Fermi sea and a change of exponents
$x=x(h)$ and Fermi momentum $k_F=k_F(h)$. 
Consequentially an increase in $h$ amounts
to (i) a suppression of the SP-instability and (ii) an incommensurate 
lattice distortion (at least for sufficiently strong fields depending
on the balance of energy gain in the commensurate versus incommensurate
case). These qualitative features are observed 
in \refto{KirKei96,Lorenz97}.
Let us consider in some detail the response of the magnetic
system (for small magnetization $m$) to a sinusoidal 
modulation of the exchange couplings (with wavevector $2k_F$). 
We are interested in the effects of Umklapp processes which favour
commensurate modulations.
For $m=0$ ($2k_F=\pi$) we expect $E_{\rm mag}=
-c_1\delta^\alpha$ and for $m\not= 0$ ($2k_F\not=\pi$) only $E_{\rm mag}=
-2 c_1(\delta/2)^\alpha=-2^{1-\alpha} c_1\delta^\alpha$. Hence, the groundstate
energy of the total system as a function
of the magnetization $m$ shows a jump at $m=0$. This is the origin of a
1st order transition from the commensurate to the incommensurate 
phase at a finite critical field $h_c$. Whether higher harmonics 
(``solitonic modulations'') can actually change the order of the transition
from 1 to 2 cannot be answered decisively in this approach. 

Lastly, we want to comment on logarithmic corrections to the above
results. These corrections typically appear if marginal
operators exist, i.e. if there are interaction terms compatible with the
usual symmetries of the lattice system (lattice translation, spin
rotation), not however with strict conformal invariance in the continuum
limit. Furthermore, these interactions have to possess the scaling dimension
$x_m=2$. From the required symmetry properties only the case $(S,d)=(0,2)$
in (\ref{scaldim}) is singled out with scaling dimension $4\xi(K)^2=4x_e$. 
With a glance to Fig.\ref{fig:fig1} we directly see that logarithmic
corrections only occur for vanishing external field. For completeness
we want to note the physical scenario at this point. The function
$\delta(L)$ in (\ref{finitesize}), which typically decays algebraically
now has a leading logarithmic contribution $\delta(L)\simeq b/\log L$.
The amplitude $b$ also determines the {\em multiplicative} logarithmic corrections
\refto{AfflGSZ89} to \hfill\break
(i) the two-point correlation function of an operator $o(r)$
\begin{equation}
\langle o(0)o(r)\rangle\simeq\frac{1}{[r(\log r)^b]^{2x}}
\end{equation}
(ii) the response of the critical system to a (relevant) perturbation 
with $H'=\sum_{r=1}^Lo(r)$ opening a gap
\begin{equation}
m_0\simeq\left(\frac{\delta}{(\log \delta)^{bx}}\right)^\frac{1}{2-x},
\end{equation}
where in both cases $x$ is the scaling dimension of the operator $o(r)$
and $\delta$ is the (small) coupling parameter.
In our case and vanishing external field we have $x=1/2$ and
$b=3/2$ \refto{AfflGSZ89} such that
$m_0\simeq\delta^{2/3}/(\log\delta)^{1/2}$.

\section{Conclusion}

In this work two aspects of the critical spin-1/2 Heisenberg chain
were treated. First, the logarithmic corrections to the leading
low-temperature properties in the specific heat and susceptibility
have been presented. For this purpose we have developed a systematic
expansion of the non-linear integral equation governing the thermodynamics
of the system. In physical terms, the low-temperature singularities 
are caused by the algebraic asymptotics of the $S$-matrix of elementary 
excitations in dependence on the difference in the spectral parameters.
This is consistent with the absence of logarithmic corrections for
models with long-range interactions \refto{KuramotoK95}.
In addition, numerical results have been presented for the Heisenberg
chain down to very low temperatures. 

Second, at zero temperature we have treated the perturbation of the 
Heisenberg chain by inhomogeneous couplings. Quite generally this problem
can be solved by use of scaling relations. In our case the Spin-Peierls
exponents in presence of an external magnetic field have been determined.
We like to note that a similar reasoning is applicable to many questions
about critical quantum systems perturbed by additional couplings, e.g.
interchain couplings.

In future work we want to combine the two approaches, calculation of
finite temperature correlation lengths and exact scaling dimensions, to
a more detailed description of the spin-Peierls scenario of the Heisenberg
system, e.g the critical lines of the phase diagram.

\section*{Acknowledgements}
The author acknowledges valuable discussions with K. Fabricius.
The comparison to his numerical data for the thermodynamics of finite systems 
proved essential in achieving high accuracy in the treatment of 
the non-linear integral equations. The author is grateful to 
Wuppertal University where parts of the work were performed.
Furthermore, financial   support is  acknowledged by the   {\it  Deutsche
Forschungsgemeinschaft} under grant  No.   Kl~645/3-1 and support by
the research program of the 
Sonderforschungsbereich 341, K\"oln-Aachen-J\"ulich. 

\newpage
\def\and{and\ }
\def\eds{eds.\ }
\def\edi{ed.\ }
\references

\def\mtb{M. T. Batchelor}
\def\rjb{R. J. Baxter}
\def\dk{D. Kim}
\def\pap{P. A. Pearce}
\def\nyr{N. Yu. Reshetikhin}
\def\ak{A. Kl\"umper}

\refis{AbramS64} M. Abramowitz, I. A. Stegun, ``Handbook of 
Mathematical Functions", Washington, U.S. National Bureau of Standards 1964;
New York, Dover 1965.

\refis{Affl86} I. Affleck, \prl 56, 746, 1986

\refis{AKLT87} I. Affleck, T. Kennedy, E. H. Lieb \and H. Tasaki, \prl 59, 
799, 1987

\refis{AKLT88} I. Affleck, T. Kennedy, E. H. Lieb \and H. Tasaki, \cmp 115, 
477, 1988

\refis{AfflGSZ89} I. Affleck, D. Gepner, H. J. Schulz \and T. Ziman,
\jpa 22, 511, 1989

\refis{AfflM88} I. Affleck \and J.B. Marston, \jpc 21, 2511, 1988

\refis{AkutDW89} Y. Akutsu, T. Deguchi \and M. Wadati, in Braid Group, Knot
Theory and Statistical
Mechanics, \eds C. N. Yang \and M. L. Ge, World Scientific, Singapore, 1989

\refis{AkutKW86a} Y. Akutsu, A. Kuniba \and M. Wadati,\jpj 55, 1466, 1986

\refis{AkutKW86b} Y. Akutsu, A. Kuniba \and M. Wadati,\jpj 55, 2907, 1986

\refis{AlcaBB87} F. C. Alcaraz, M. N. Barber \and \mtb,\prl 58, 771, 1987

\refis{AlcaBB88} F. C. Alcaraz, M. N. Barber \and \mtb,\annp 182, 280, 1988

\refis{AlcaBGR88} F. C. Alcaraz, M. Baake, U. Grimm \and V. Rittenberg,
\jpa 21, L117, 1988

\refis{AlcaM89}  F. C. Alcaraz \and M. J. Martins, \jpa 22, 1829, 1989

\refis{AlcaM90}  F. C. Alcaraz \and M. J. Martins, \jpa 23, 1439-51, 1990

\refis{Alex75} S. Alexander, \pla 54, 353-4, 1975

\refis{AndrBF84} G. E. Andrews, \rjb\ \and P. J. Forrester, \jsp 35, 193,
1984

\refis{And87} P. W. Anderson, Science 235, 1196, 1987

\refis{And90} P. W. Anderson, \prl 64, 1839, 1990

\refis{ArrAl94} L. Arrachea \and A. A. Aligia, \prl 73, 2240, 1994

\refis{BDV82} O. Babelon, H. J. de Vega, \and C. M. Viallet, \npb 200 [FS4], 266, 1982

\refis{BarmaS77} M. Barma \and S. Shastry, \pla 61, 15, 1977

\refis{BarbBP87} \mtb, M.N. Barber \and \pap,\jsp 49, 1117, 1987

\refis{BarbB89} M.N. Barber \and \mtb, \prb 40, 4621, 1989

\refis{Barb91} M.N. Barber, \physica A 170, 221, 1991

\refis{BaresB90} P. A. Bares \and G. Blatter, \prl 64, 2567, 1990

\refis{BaresBO90} P. A. Bares, G. Blatter \and M. Ogata,
\prb 44, 130, 1991

\refis{Bariev8182} R. Z. Bariev, \tmp 49, 261, 1981; 1021, 1982

\refis{Bariev82} R. Z. Bariev, \tmp 49, 1021, 1982

\refis{Bariev91} R. Z. Bariev, \jpa 24, L549, 1991; L919, 1991

\refis{BarievKSZ93} R. Z. Bariev, A. Kl\"{u}mper, A. Schadschneider 
\and J. Zittartz, \jpa 26, 1249, 1993; 4863

\refis{BarievKSZ93sum} R. Z. Bariev, A. Kl\"{u}mper, A. Schadschneider 
\and J. Zittartz, \jpa 26, 1249, 1993; 4863; \physica B 194-196, 1417, 1994

\refis{BarievKSZ94b} R. Z. Bariev, A. Kl\"{u}mper, A. Schadschneider 
\and J. Zittartz, \prb 50, 9676, 1994

\refis{BarievKSZ94a} R. Z. Bariev, A. Kl\"{u}mper, A. Schadschneider 
\and J. Zittartz, \zpb 96, 395, 1995

\refis{BarievKSZmult} R. Z. Bariev, A. Kl\"{u}mper, A. Schadschneider 
\and J. Zittartz, \prb 50, 9676, 1994; \jpa 28, 2437, 1995

\refis{ZittKSB94} J. Zittartz, A. Kl\"{u}mper, A. Schadschneider 
\and R. Z. Bariev, \physica B 194-196, 1417, 1994

\refis{Bariev94a} R. Z. Bariev, \prb 49, 1474, 1994

\refis{Bariev94b} R. Z. Bariev, \jpa 27, 3381, 1994

\refis{BarievKZ95} R. Z. Bariev, A. Kl\"{u}mper, 
\and J. Zittartz, \eurolett 32, 85, 1995

\refis{BarouchM71} E. Barouch \and B. M. McCoy, \pra 3, 786, 1971

\refis{Baxt70} \rjb,\jmp 11, 3116, 1970

\refis{Baxt71b} \rjb,\prl 26, 834, 1971

\refis{Baxt72} \rjb,\annp 70, 193, 1972

\refis{Baxt73} \rjb,\jsp 8, 25, 1973

\refis{Baxt80} \rjb,\jpa 13, L61--70, 1980

\refis{Baxt81a} \rjb,\physica 106A, 18--27, 1981

\refis{Baxt81b} \rjb,\jsp 26, 427--52, 1981

\refis{Baxt82a} \rjb,\jsp 28, 1, 1982

\refis{Baxt82b} \rjb, ``Exactly Solved Models in Statistical Mechanics",
Academic Press, London, 1982.

\refis{BaxtP82} \rjb\space \and \pap,\jpa 15, 897, 1982

\refis{BaxtP83} \rjb\space \and \pap,\jpa 16, 2239, 1983

\refis{BazhR89} V.V. Bazhanov \and \nyr,\ijmpa 4, 115--42, 1989 

\refis{BazhB93} V.V. Bazhanov \and \rjb,\physica A 194, 390--396, 1993 

\refis{BednorzM86} J. G. Bednorz \and K. A. M\"uller, \zpb 64, 189, 1986

\refis{Beduerf95} G. Bed\"urftig \and H. Frahm, \jpa 28, 4453, 1995

\refis{BelaPZ84} A. A. Belavin, A. M. Polyakov \and A. B. Zamolodchikov,
\npb 241, 333, 1984

\refis{Bethe31} H. A. Bethe,\zp 71, 205, 1931

\refis{BlotCN86} H. W. J. Bl\"ote, J. L. Cardy \and M. P. Nightingale, \prl
56, 742,
1986

\refis{BogK89} N. M. Bogoliubov \and V. E. Korepin, \ijmpb 3, 427-439, 1989

\refis{BoerKS95} J. de Boer, V. E. Korepin, A. Schadschneider, \prl 74,
789, 1995

\refis{BogIR86} N. M. Bogoliubov, A.\ G.\ Izergin \and 
N.\ Y.\ Reshetikhin, \jetpl 44, 405, 1986

\refis{Brack94} A. J. Bracken, M. D. Gould, J. R. Links \and Y.-Z. Zhang,
\prl 74, 2768, 1995

\refis{BretzD71} M. Bretz \and J. G. Dash, \prl 27, 647, 1971

\refis{Bretz77} M. Bretz, \prl 38, 501, 1977

\refis{Buy86} W. J. L. Buyers, R. M. Morra, R. L. Armstrong, P. Gerlach
\and K. Hirakawa, \prl 56, 371, 1986

\refis{EssKS9293} F. H. L. Essler, V. E. Korepin \and K.Schoutens, 
\prl 68, 2960, 1992; \prl 70, 73, 1993

\refis{EssK94} F. H. L. Essler \and V. E. Korepin,
\prl 72, 908, 1994

\refis{KawUO89} N. Kawakami, T. Usuki \and A. Okiji, \pla 137, 287, 1989

\refis{UsukiKO90} T. Usuki, N. Kawakami \and A. Okiji, \jpj 59, 1357, 1990

\refis{KawY90} N. Kawakami \and S.-K. Yang, \prl 65, 2309, 1990

\refis{KawY91} N. Kawakami \and S.-K. Yang, \prb 44, 7844, 1991

\refis{Kaw93} N. Kawakami, \prb 47, 2928, 1993

\refis{KorBI93} V. E. Korepin, N.M. Bogoliubov, \and A.G. Izergin,
``Quantum Inverse Scattering Method and Correlation
Functions", Cambridge University Press, 1993.

\refis{Morra88} R. M. Morra, W. J. L. Buyers, R. L. Armstrong \and K. Hirakawa,
\prb 38, 543, 1988

\refis{SimAl93} M. E. Simon \and A. A. Aligia, \prb 48, 7471, 1993

\refis{ShasS90} B. S. Shastry \and B. Sutherland, \prl 66, 243, 1990

\refis{Shastry88} B. S. Shastry, \jsp 50, 57, 1988

\refis{Shiba72} H. Shiba, \prb 6, 930, 1972

\refis{Stei87} M. Steiner, K. Kakurai, J. K. Kjems, D. Petitgrand \and R. Pynn,
\jappp 61, 3953, 1987

\refis{Tun90} Z. Tun, W. J. L. Buyers, R. L. Armstrong, K. Hirakawa \and
B. Briat, \prb 42, 4677, 1990

\refis{Tun91} Z. Tun, W. J. L. Buyers, A. Harrison \and J. A. Rayne, \prb 43,
13331, 1991

\refis{Ren87} J. P. Renard, M. Verdaguer, L. P. Regnault, W. A. C. Erkelens,
J. Rossa-Mignod \and W. G. Stirling, \eurolett 3, 945, 1987

\refis{Ren88} J. P. Renard, M. Verdaguer, L. P. Regnault, W. A. C. Erkelens,
J. Rossa-Mignod, J. Ribas, W. G. Stirling \and C. Vettier, \jappp 63, 3538, 1988

\refis{Reg89} L. P. Regnault, J. Rossa-Mignod, J. P. Renard, M. Verdaguer
\and C. Vettier, \physica B 156 \& 157, 247, 1989

\refis{Colom87} P. Colombet, S. Lee, G. Ouvrard \and R. Brec, \jcr, 134, 1987

\refis{Coll74} C. F. Coll, \prb 9, 2150, 1974

\refis{deGroot82} H. J. M. de Groot, L. J . de Jongh, R. D. Willet \and
J. Reeyk, \jappp 53, 8038, 1982

\refis{Capp88} A. Cappelli, Recent Results in Two-Dimensional Conformal
Field
Theory, in Proceedings of the XXIV International Conference on High Energy
Physics,
\eds R. Kotthaus \and J. K\"uhn, Springer, Berlin, 1988

\refis{CappIZ87a} A. Cappelli, C. Itzykson \and J.-B. Zuber, \npb {280
[FS18]},
445--65, 1987

\refis{CappIZ87b} A. Cappelli, C. Itzykson \and J.-B. Zuber, \cmp 113,
1--26, 1987

\refis{Card84a} J. L. Cardy, \jpa 17, L385, 1984

\refis{Card86a} J. L. Cardy, \npb {270 [FS16]}, 186, 1986

\refis{Card86b} J. L. Cardy, \npb {275 [FS17]}, 200, 1986

\refis{Card88} J. L. Cardy, ``Phase Transitions and Critical Phenomena,
Vol.11",
\eds C. Domb \and J.L. Lebowitz, Academic Press, London 1988

\refis{Card89} J. L. Cardy, Conformal Invariance and Statistical Mechanics,
in Les
Houches, Session XLIV, Fields, Strings and Critical Phenomena, \eds E.
Br\'ezin \and
J. Zinn-Justin, 1989

\refis{ChoiKK90} J.-Y. Choi, K. Kwon \and D. Kim, \eurolett xx, to appear,
1990

\refis{ChoiKP89} J.-Y. Choi, D. Kim \and \pap, \jpa 22, 1661--71, 1989

\refis{CvetDS80} D. M. Cvetkovic, M. Doob \and H. Sachs, ``Spectra of
Graphs", Academic Press, London 1980

\refis{DateJKMO87} E. Date, M. Jimbo, A. Kuniba, T. Miwa, \and M. Okado,
\npb
B290, 231--273, 1987

\refis{DateJKMO88} E. Date, M. Jimbo, A. Kuniba, T. Miwa, \and M. Okado,
\aspm 16,
17, 1988

\refis{DateJMO86} E. Date, M. Jimbo, T. Miwa \and M. Okado,\lmp 12, 209,
1986

\refis{DateJMO87} E. Date, M. Jimbo, T. Miwa \and M. Okado,\prb 35, 2105--7,
1987

\refis{DaviP90} B. Davies \and \pap, \ijmpb {}, this issue, 1990

\refis{deVeK87} H. J. de Vega \and M. Karowski, \npb {285 [FS19]}, 619, 1987

\refis{deVeW85} H. J. de Vega \and F. Woynarovich,\npb 251, 439, 1985

\refis{deVeW90} H. J. de Vega \and F. Woynarovich,\jpa 23, 1613, 1990

\refis{DestdeVeW92} C. Destri \and H. J. de Vega,\prl 69, 2313, 1992

\refis{DestdeVeW95} C. Destri \and H. J. de Vega, \npb 438, 413, 1995

\refis{diFrSZ87} P. di Francesco, H. Saleur \and J.-B. Zuber, \jsp 49,
57--79, 1987

\refis{diFrZ89} P. di Francesco \and J.-B. Zuber, $SU(N)$ Lattice Models
Associated
with Graphs, Saclay preprint SPhT/89-92, 1989

\refis{DijkVV88} R. Dijkgraaf, E. Verlinde \and H. Verlinde, in Proceedings
of the
1987 Copenhagen Conference, World Scientific, 1988

\refis{DijkVVV89} R. Dijkgraaf, C. Vafa, E. Verlinde \and H. Verlinde,\cmp
123, 485, 1989

\refis{DombG76} ``Phase Transitions and Critical Phenomena, Vol.6",
Academic Press, London 1976

\refis{FateZ85} V. A. Fateev \and A. B. Zamolodchikov, \jetp 62, 215, 1985

\refis{FendG89} P. Fendley \and P. Ginsparg, \npb 324, 549--80, 1989

\refis{FodaN89} O. Foda \and B. Nienhuis, \npb {},{},1989

\refis{FoersK93} A. Foerster \and M. Karowski, \npb 408 [FS], 512, 1993

\refis{FrieQS84} D. Friedan, Z. Qiu \and S. Shenker, \prl 52, 1575, 1984; in
``Vertex Operators in Mathematics and Physics", \eds J. Lepowsky, S.
Mandelstam \and
I.M. Singer, Springer, 1984

\refis{FrahmK90} H. Frahm \and V. E. Korepin, \prb 42, 10553, 1990

\refis{FrahmYF90} H. Frahm, N.-C. Yu \and M. Fowler, \npb 336, 396, 1990

\refis{Frei93} W.-D. Freitag, Dissertation, Universit\"at zu K\"oln, 1993

\refis{Gaudin83} M. Gaudin, ``La Fonction d'Onde de Bethe", Masson, Paris, 1983.

\refis{GepnQ87} D. Gepner \and Z. Qiu, \npb 285, 423--53, 1987

\refis{Gins88} P. Ginsparg,\npb {295 [FS21]}, 153--70, 1988

\refis{Gins89a} P. Ginsparg, Applied Conformal Field Theory, in Les
Houches,
Session XLIV, Fields, Strings and Critical Phenomena, \eds E. Br\'ezin \and
J.
Zinn-Justin, 1989

\refis{Gins89b} P. Ginsparg, Some Statistical Mechanical Models and
Conformal Field
Theories, Trieste Spring School Lectures, HUTP-89/A027

\refis{GradR80} I.S. Gradshteyn \and I.M. Ryzhik, ``Tables of Integrals,
Series and Products", Academic
Press, New York, 1980.

\refis{Grif72} R. B. Griffiths, ``Phase Transitions and Critical Phenomena,
Vol.1",\eds C. Domb \and M. S. Green, Academic Press, London 1972

\refis{Hald83a} F. D. M. Haldane, \prl 50, 1153, 1983

\refis{Hald83b} F. D. M. Haldane, \pla 93, 464, 1983

\refis{Hame85} C. J. Hamer,\jpa 18, L1133, 1985

\refis{Hame86} C. J. Hamer,\jpa 19, 3335, 1986

\refis{Hirsch89a} J. E. Hirsch, \pla 134, 451, 1989

\refis{Hirsch89b} J. E. Hirsch, \physica  C 158, 326, 1989

\refis{Hubbard} J. Hubbard, \prs 276, 238, 1963

\refis{HuiD93} A. Hui \and S. Doniach, \pr 48, 2063, 1993

\refis{Huse82} D. A. Huse,\prl 49, 1121--4, 1982

\refis{Huse84} D. A. Huse, \prb 30, 3908, 1984

\refis{Idz94} M. Idzumi, T. Tokihiro \and M. Arai, \jpI 4, 1151, 1994

\refis{ItzySZ88} C. Itzykson, H. Saleur \and J-B. Zuber, ``Conformal
Invariance and Applications to
Statistical Mechanics", World Scientific, Singapore, 1988

\refis{JapaMu94} G. Japaridze \and E. M\"uller-Hartmann , \Annp 3, 163, 1994

\refis{JimbM84} M. Jimbo \and T. Miwa, \aspm 4, 97--119, 1984

\refis{JimbMO87} M. Jimbo, T. Miwa \and M. Okado, \lmp 14, 123--31, 1987

\refis{JimbMO88} M. Jimbo, T. Miwa \and M. Okado, \cmp 116, 507--25, 1988

\refis{JimbMT89} M. Jimbo, T. Miwa \and A. Tsuchiya,``Integrable Systems in
Quantum Field Theory and
Statistical Mechanics", \aspm 19, ,1989

\refis{JohnKM73} J.D. Johnson, S. Krinsky, \and B.M. McCoy,\pra 8, 2526,
1973

\refis{JuettK96} G. J\"uttner \and A. Kl\"umper, cond-mat/9606192

\refis{JuettKS96} G. J\"uttner, A. Kl\"umper, \and J. Suzuki, Preprint 1996

\refis{Kac79} V. G. Kac, \lnp 94, 441--445, 1979

\refis{KadaB79} L. P. Kadanoff \and A. C. Brown, \annp 121, 318--42, 1979

\refis{Karo88} M. Karowski, \npb {300 [FS22]}, 473, 1988

\refis{Karn94} I. N. Karnaukhov, \prl 73, 1130, 1994

\refis{Karn95} I. N. Karnaukhov, \prb 51, 7858, 1995

\refis{Kato87} A. Kato, \mpla 2, 585, 1987

\refis{KimP87} \dk\space \and \pap,\jpa 20, L451--6, 1987

\refis{KimP89}  \dk\space \and \pap,\jpa 22, 1439--50, 1989

\refis{Kiri89} E. B. Kiritsis, \plb  217, 427, 1989

\refis{KiriR86} A. N. Kirillov \and N. Yu. Reshetikhin,\jpa 19, 565, 1986

\refis{KiriR87} A. N. Kirillov \and N. Yu. Reshetikhin,\jpa 20, 1565, 1987

\refis{KlassM90} T. R. Klassen \and E. Melzer, \npb 338, 485, 1990

\refis{KlassM91} T. R. Klassen \and E. Melzer, \npb 350, 635, 1991

\refis{KlumBiq} A. Kl\"{u}mper, \eurolett 9, 815, 1989; \jpa 23, 809, 1990

\refis{KlumB90} A. Kl\"{u}mper \and \mtb,\jpa 23, L189, 1990

\refis{KlumBP91} A. Kl\"{u}mper, \mtb \ \and \pap, \jpa 24, 3111--3133, 1991

\refis{KlumP91} A. Kl\"{u}mper \and \pap, \jsp 64, 13--76, 1991

\refis{Klum92c} A. Kl\"{u}mper, unver"offentlichte Rechnungen, (1992)

\refis{KlumZ88} A. Kl\"{u}mper \and J. Zittartz,\zpb 71, 495, 1988

\refis{KlumZ88App} A. Kl\"{u}mper \and J. Zittartz,\zpb 71, 495, 1988, 
Appendix A

\refis{KlumZ89} A. Kl\"{u}mper \and J. Zittartz,\zpb 75, 371, 1989

\refis{KlumZ8VM} A. Kl\"{u}mper \and J. Zittartz,\zpb 71, 495, 1988;
\zpb 75, 371, 1989

\refis{KlumSZ89} A. Kl\"{u}mper, A. Schadschneider \and J. Zittartz,
\zpb 76, 247, 1989

\refis{KlumSZ90} A. Kl\"{u}mper, A. Schadschneider \and J. Zittartz,
\zpb 78, 99, 1990

\refis{KlumSZMPG} A. Kl\"{u}mper, A. Schadschneider \and J. Zittartz,
\jpa 24, L955-L959, 1991; \zpb 87, 281-287, 1992

\refis{Klum89} \ak, \eurolett 9, 815, 1989

\refis{KlumP92} \ak\  \and \pap, \physica 183A, 304-350, 1992

\refis{Klum92} \ak , \Annp 1, 540, 1992

\refis{Klum93} \ak , \zpb 91, 507, 1993

\refis{KlumTH} \ak , \Annp 1, 540, 1992; \zpb 91, 507, 1993

\refis{Klum92b} \ak, in preparation

\refis{KlumWZ93} \ak, T. Wehner \and J. Zittartz, \jpa 26, 2815, 1993

\refis{Klum94} \ak , in Vorbereitung

\refis{KlumWeh94} \ak\ \and T. Wehner, in Vorbereitung

\refis{KlumWeh95} \ak\ \and T. Wehner, in preparation

\refis{KlumBar95} \ak\ \and R. Z. Bariev, \npb 458, 623, 1995

\refis{Knabe88} S. Knabe, \jsp 52, 627, 1988

\refis{Koma} T. Koma, \ptp 78, 1213, 1987; \bf 81, \rm 783, (1989)

\refis{KorepinS90} V. E. Korepin \and N. A. Slavnov, \npb 340, 759, 1990

\refis{ItsIK92} A. R. Its, A. G. Izergin \and V. E. Korepin, \physica D 54, 351, 1992

\refis{ItsIKS93} A. R. Its, A. G. Izergin, V. E. Korepin \and N. A. Slavnov, 
\prl 70, 1704, 1993

\refis{IKR89} A. G. Izergin, V. E. Korepin \and N. Yu. Reshetikhin, \jpa 22, 2615, 1989

\refis{KuniY88} A. Kuniba \and T. Yajima,\jsp 52, 829, 1988

\refis{KuliRS81} P. P. Kulish, N. Yu. Reshetikhin \and E. K. Sklyanin, \lmp
5, 393, 1981 

\refis{Kuniba92} A. Kuniba, ``Thermodynamics of the $U_q(X_r^{(1)})$ Bethe Ansatz System with $q$ a Root of Unity", ANU preprint (1991)    

\refis{LeeS88} K. Lee \and P. Schlottmann, \jpcoll 49 C8, 709, 1988

\refis{Lewi58} L. Lewin, Dilogarithms and Associated Functions, MacDonald,
London, 1958

\refis{LiebWu68} E. H. Lieb \and F. Y. Wu, \prl 20, 1445, 1968

\refis{LiebWu72} E. H. Lieb \and F. Y. Wu, 
``Phase Transitions and Critical Phenomena,
Vol.1",
\eds C. Domb \and M. S. Green, Academic Press, London 1988

\refis{LinH95} H. Q. Lin \and J. E. Hirsch, \prb 52, 16155, 1995

\refis{LutherP74} A. Luther \and I. Peschel, \prb 9, 2911, 1974

\refis{Martins91} M. J. Martins, \prl 22, 419, 1991 and private communication 
(1991)

\refis{Mizuta95} Mizuta, T. Nagao and M. Wadati,
J. Phys. Soc. Japan (1995).

\refis{Muell} E. M\"uller-Hartmann, unpublished results, (1989)

\refis{Muell89} E. M\"uller-Hartmann, unver"offentlichte Ergebnisse, (1989)

\refis{Mura89} J. Murakami, \aspm 19, 399--415, 1989

\refis{Nien87} B. Nienhuis, in Phase Transitions and Critical Phenomena,
Vol.11,
\eds C. Domb \and J.L. Lebowitz, Academic Press, 1987

\refis{NahmRT92} W. Nahm, A. Recknagel \and M. Terhoven, Preprint ``Dilogarithm
identities in conformal field theory'', 1992

\refis{NighB86} M. P. Nightingale \and H. W. J. Bl"ote, \prb 33, 659, 1986

\refis{Ovch70} A. A. Ovchinnikov, \jetp 30, 1160, 1970

\refis{OwczB87} A. L. Owczarek \and \rjb,\jsp 49, 1093, 1987

\refis{ParkBiq} J. B. Parkinson,\jpc 20, L1029, 1987; \jpc 21, 3793, 1988; 
\jphc 8, 1413, 1988

\refis{ParkB85} J. B. Parkinson \and J. C. Bonner, \prb 32, 4703, 1985

\refis{PaczP90} I. D. Paczek \and J. B. Parkinson,\jpcon 2, 5373, 1990

\refis{Pasq87a} V. Pasquier,\npb {285 [FS19]}, 162, 1987

\refis{Pasq87b} V. Pasquier,\jpa 20, {L217, L221}, 1987

\refis{Pasq87c} V. Pasquier,\jpa 20, {L1229, 5707}, 1987

\refis{Pasq88} V. Pasquier,\npb {B295 [FS21]}, 491--510, 1988

\refis{Pear85} \pap,\jpa 18, 3217--26, 1985

\refis{Pear87prl} \pap,\prl 58, 1502--4, 1987

\refis{Pear87jpa} \pap,\jpa 20, 6463--9, 1987

\refis{Pear90ijmpb} \pap,\ijmpb 4, 715--34, 1990

\refis{PearB90} \pap\space \and \mtb, \jsp 60, 77--135, 1990

\refis{PearK87} \pap \and \dk, \jpa, 20, 6471-85, 1987

\refis{PearS88} \pap\space \and K. A. Seaton,\prl 60, 1347, 1988

\refis{PearS89} \pap\space \and K. A. Seaton,\annp 193, 326, 1989

\refis{PearS90} \pap\space \and K. A. Seaton,\jpa 23, 1191--1206, 1990

\refis{Pear91} \pap, Row Transfer Matrix Functional Equations for
$A$--$D$--$E$ Lattice Models,
 to be published, 1991

\refis{PearK91} \pap\space \and A. Kl\"umper, \prl 66, 974, 1991

\refis{Pear92} \pap, \ijmpa 7, Suppl.1B, 791, 1992

\refis{PerkS81} J. H. H. Perk \and C. L. Schultz, \pla 84, 407, 1981

\refis{PensK86} K. Penson \and M. Kolb, \prb 33, 1663, 1986; M. Kolb \and K.
Penson, \jsp 44, 129, 1986

\refis{PfannFr96} M. P. Pfannm\"uller \and H. Frahm, cond-mat/9604082

\refis{RamosM96} P. B. Ramos \and M. J. Martins, hep-th/9604072

\refis{Resh83jetp} \nyr,\jetp 57, 691, 1983

\refis{Resh83lmp} \nyr, \lmp 7, 205--13, 1983

\refis{Sale88} H. Saleur, Lattice Models and Conformal Field Theories, in
Carg\`ese
School on Common Trends in Condensed Matter and Particle Physics, 1988

\refis{SaleB89} H. Saleur \and M. Bauer, \npb 320, 591--624, 1989

\refis{SaleD87} H. Saleur \and P. di Francesco, Two Dimensional Critical
Models on a
Torus, in Brasov Summer School on Conformal Invariance and String Theory,
1987

\refis{Samuel73} E. J. Samuelson, \prl 31, 936, 1973

\refis{Schlott87} P. Schlottmann, \prb 36, 5177, 1987

\refis{Sch87} P. Schlottmann, \jpc 4, 7565, 1987

\refis{Schlott92} P. Schlottmann, \jpc 4, 7565, 1992

\refis{Schul83} C. L. Schultz, \physica 122A, 71, 1983

\refis{SeatP89} K. A. Seaton \and \pap\space, \jpa 22, 2567--76, 1989

\refis{Strog79} Yu. G. Stroganov, \pla 74, 116, 1979

\refis{Suth70} B. Sutherland, \jmp 11, 3183, 1970

\refis{Suth75} B. Sutherland, \prb 12, 3795, 1975

\refis{Suth} B. Sutherland, in Exactly Solved Problems in Condensed Matter and
Relativistic Field Theory (Lecture Notes in Physics 242), \eds 
B. S. Shastry, S. S. Iha \and V. Sigh, Springer, Berlin, 1982

\refis{Suzuki85} M. Suzuki, \prb 31, 2957, 1985

\refis{SuzukiI87} M. Suzuki \and M. Inoue, \ptp 78, 787, 1987

\refis{Suzuki87} M. Suzuki, in ``Quantum Monte Carlo Methods in 
Equilibrium and Nonequilibrium Systems",
\edi M. Suzuki, Springer Verlag, 1987

\refis{SuzukiAW90} J. Suzuki, Y. Akutsu \and M. Wadati, \jpj 59, 2667, 1990

\refis{SuzukiNW92} J. Suzuki, T. Nagao \and M. Wadati, \ijmpb 6, 1119, 1992

\refis{Tak69} M. Takahashi, \ptp 42, 1098, 1969

\refis{Tak71} M. Takahashi, \ptp 46, 401, 1971

\refis{Tak72} M. Takahashi, \ptp 47, 69, 1972

\refis{Tak74} M. Takahashi, \ptp 52, 103, 1974

\refis{TakTBA} M. Takahashi, \ptp 46, 401, 1971; \ptp 50, 1519, 1973

\refis{Tak91} M. Takahashi, \prb 43, 5788, 1991; \prb 44, 12382, 1991

\refis{Tak91a} M. Takahashi, \prb 43, 5788, 1991

\refis{Tak91b} M. Takahashi, \prb 44, 12382, 1991

\refis{TempL71} H. N. V. Temperley \and E. H. Lieb, \prs 322, 251, 1971

\refis{Tetel82} M. G. Tetel'man, \jetp 55, 306, 1982

\refis{TruS83} T. T. Truong \and K. D. Schotte, \npb 220, 77, 1983

\refis{Tsun91} H. Tsunetsugu, \jpj 60, 1460, 1991

\refis{vonGR87} G. von Gehlen \and V. Rittenberg, \jpa 20, 227, 1987

\refis{WadaDA89} M. Wadati, T. Deguchi \and Y. Akutsu, \prep 180, 247--332,
1989

\refis{Woyn} F. Woynarovich, \jpc 15, 85, 1982; \jpc 15, 97, 1982; 
\jpc 16, 5293, 1983; \jpc 16, 6593, 1983

\refis{Woyn83} F. Woynarovich, \jpc 16, 5293, 1983

\refis{Woyn87} F. Woynarovich, \prl 59, 259, 1987

\refis{WoynE87} F. Woynarovich \and H.-P. Eckle, \jpa 20, L97, 1987

\refis{Woyn89} F. Woynarovich, \jpa 22, 4243, 1989

\refis{Yang69} C. N. Yang \and C. P. Yang, \jmp 10, 1115, 1969

\refis{Yang66} C. N. Yang \and C. P. Yang, \pr 147, 303, 1966; 150, 321

\refis{Yang62} C. N. Yang, \rmp 34, 691, 1962

\refis{Yang67} C. N. Yang, \prl 19, 1312, 1967

\refis{CPYang67} C. P. Yang, \prl 19, 586, 1967

\refis{YangG89} C. N. Yang \and M. L. Ge (Editors), Braid Group, Knot Theory
and Statistical
Mechanics, World Scientific, Singapore, 1989

\refis{ZamoF80} A. B. Zamolodchikov \and V. Fateev, \sjnp 32, 198, 1980

\refis{Zamo80} A. B. Zamolodchikov, \jetp 52, 325, 1980; \cmp 79, 489, 1981

\refis{Zamo91} Al. B. Zamolodchikov, \plb 253, 391--4, 1991; \npb 358, 
497--523, 1991

\refis{Zamo91a} Al. B. Zamolodchikov, \plb 253, 391--4, 1991

\refis{Zamo91b} Al. B. Zamolodchikov, \npb 358, 497--523, 1991

\refis{ZhangR88} F. C. Zhang and T. M. Rice, \prb 37, 3759, 1988

\refis{FadTak81}
L.~Faddeev and L.~Takhtajan, Phys. Lett. A 85, 375, 1981

\refis{OkaNom92} K. Okamoto and K. Nomura, \pla 169, 433, 1992

\refis{NomOka93} K. Nomura and K. Okamoto, \jpsj 62, 1123, 1993

\refis{Nom93} K. Nomura, \prb 48, 16814, 1993

\refis{EggertAT94} S. Eggert, I. Affleck and M. Takahashi, \prl 73, 332, 1994

\refis{KawanoT95} K. Kawano and M. Takahashi, \jpsj 64, 4331, 1995

\refis{KlumTak98} \ak \ and M. Takahashi, to be published

\refis{CrossFisher79} M.C. Cross and D.S. Fisher, \prb 19, 402, 1979

\refis{Cross79} M.C. Cross, \prb 20, 4606, 1979

\refis{Arai96} M. Arai et al., \prl 77, 3649, 1996

\refis{Takagi96} S. Takagi, H. Deguchi, K. Takeda, M. Mito
and M. Takahashi, \jpsj 65, 1934, 1996

\refis{Karbach} M. Karbach and K.-H. M\"utter, \jpa 28, 4469 , 1995

\refis{Peierls} R. E. Peierls, ``Quantum Theory of Solids'', Oxford 
University, London, 1955

\refis{KuramotoK95} Y. Kuramoto and Y. Kato, \jpsj 64, 4518, 1995

\refis{Takigawa96} M. Takigawa, N. Motoyama, H. Eisaki and S. Uchida,
\prl 76, 4612, 1996

\refis{Boeskeetal97} T. B\"oske, K. Maiti, O. Knauff, K. Ruck, M. S. Golden,
G. Krabbes, J. Fink, T. Osafune, N. Motoyama, H. Eisaki and S. Uchida,
preprint, cond-mat/9709215, 1997

\refis{FleddGMSK96} A Fledderjohann, C Gerhardt, K-H Mutter, 
A Schmitt, M Karbach, \prb 54, 7168, 1996

\refis{KirKei96} V. Kiryukhin, B. Keimer, \prb 52, R704, 1996

\refis{LAZBRD96} T. Lorenz, U. Ammerahl, R. Ziemes, B. Buechner, 
A.Revcolevschi, G. Dhalenne, Phys. Rev. B 54, 15610, 1996

\refis{Lorenz97} T. Lorenz, U. Ammerahl, T. Auweiler, B. B\"uchner,
A. Revcolevschi, G Dhalenne, \prb 55, 5914, 1997

\refis{BeNaRon97} S.M. Bhattacharjee, T. Nattermann, and C. Ronnewinkel,
cond-mat/9711094

\refis{Liu95} X. Liu, J. Wosnitza, H.v. L\"ohneysen, and R.K. Kremer,
Z. Phys\. B 98, 163, 1995

\refis{Weiden95} M. Weiden, J. K\"ohler, G. Sparn, M. K\"oppen, M.
Lang, C. Geibel, and F. Steglich, Z. Phys. B 98, 167, 1995

\endreferences

\newpage

\end{document}
\bye